\documentclass{JHEP3}

\usepackage{latexsym,amsmath,amssymb}
\pdfoutput=1
\usepackage{graphicx}
\usepackage{eufrak}
\usepackage{slashed}
\usepackage{bm}
\usepackage{epsfig}
\newcommand{\exclude}[1]{}
\usepackage{epsfig}
\usepackage{dcolumn}

\newcommand{\beq}{\begin{equation}}
\newcommand{\eeq}{\end{equation}}
\newcommand{\be}{\begin{eqnarray}}
\newcommand{\ee}{\end{eqnarray}}

\newcommand{\rar}{\rightarrow}
\newcommand{\Rar}{\Rightarrow}

\def\dd{ \,\mathrm{d} }

\def\+{\dagger}
\def\la{\langle}
\def\ra{\rangle}
\def\<{\langle}
\def\>{\rangle}

\def\atop{\frac{ \alpha_{s}}{8 \pi} G_{\mu \nu}^{a}
 \tilde{G}^{\mu \nu a} }

\newcommand{\Lqcd}{\Lambda_{\mathrm{QCD}}}

\title{The QCD nature of Dark Energy}

\author{Federico R. Urban and Ariel R. Zhitnitsky \\ Department of Physics \& Astronomy, University of British Columbia, Vancouver, B.C. V6T 1Z1, Canada}

\date{\today}

\abstract{The origin of the observed dark energy could be explained entirely within the standard model, with no new fields required.  We show how the low-energy sector of the chiral QCD Lagrangian, once embedded in a non-trivial spacetime, gives rise to a cosmological vacuum energy density which   can be presented entirely in terms of QCD parameters and the Hubble constant $H$ as $\rho_\Lambda \simeq H \cdot  m_q\la\bar{q}q\ra  /m_{\eta'} \sim (4.3\cdot 10^{-3} \text{eV})^4$.  In this work we focus on the dynamics of the ghost fields that are essential ingredients of the aforementioned Lagrangian.  In particular, we argue that the Veneziano ghost, being unphysical in the usual Minkowski QFT, exhibits important physical effects if the universe is expanding.  Such effects are naturally very small as they are proportional to the rate of expansion $H/ \Lqcd \sim 10^{-41}$.  The co-existence of these two drastically different scales ($\Lqcd \sim 100 $ MeV  and $H \sim 10^{-33}$ eV) is a direct consequence of the auxiliary conditions on the physical Hilbert space that are necessary to keep the theory unitary.  The exact cancellation taking place in Minkowski space due to this auxiliary condition is slightly violated when the system is upgraded to an expanding background.  Nevertheless, this ``tiny'' effect would in fact the driving force accelerating the universe today.  We also derive the time-dependent equation of state $w(t)$ for the dark energy component which tracks the dynamics of the Veneziano ghost in a FLRW universe.}

\preprint{}

\begin{document}

\section{Prelude}\label{prelude}

It has been suggested recently~\cite{our4d} that the solution of cosmological vacuum energy puzzle may not require any new field beyond the standard model.  The idea is based on the philosophy that gravitation can not be a truly fundamental interaction, but rather it must be considered as a low-energy effective quantum field theory (QFT)~\cite{Thomas:2009uh}.  In such a case, the corresponding gravitons should be treated as quasiparticles which do not feel all the microscopic degrees of freedom, but rather are sensitive to the ``relevant excitations'' only. In this framework it is quite natural to define the ``renormalised cosmological constant'' to be zero in Minkowski vacuum wherein the Einstein equations are automatically satisfied as the Ricci tensor identically vanishes. Thus, the energy-momentum tensor in combination with this ``bare cosmological constant'' must also vanish at this specific ``point of normalisation'' to satisfy the Einstein equations.  From this definition it is quite obvious that the ``renormalised energy density'' must be proportional to the deviation from Minkowski spacetime geometry.  With this definition the effective QFT of gravity has a predictive power.  In particular, it predicts the behaviour of the system in any non-trivial geometry of the spacetime. 

The first application of this proposal was the computation of the cosmological constant in a spacetime with non-trivial topological structure~\cite{our4d}.  It was shown that the cosmological constant does not vanish if our universe can be represented by a large but finite manifold with typical size $L \simeq H^{-1}$, where $H$ is the Hubble parameter.  The cosmological vacuum energy density $\rho_{\Lambda} $ in this framework is expressed in terms of QCD parameters for $N_f=2 $ light flavours as follows:
\be
\label{rhov}
\rho_{\Lambda} \simeq \frac{2N_f  |m_q\la\bar{q}q\ra  |}{m_{\eta'} L} \sim (4.3\cdot 10^{-3} \text{eV})^4 \, .
\ee
This estimate should be compared with the observational value $ \rho_{\Lambda} \approx (2.3\cdot 10^{-3} \text{eV})^4$~\cite{sn1,sn2,wmap}.  The deviation of the cosmological constant from zero is entirely due to the large but finite size $L$ of the manifold.

The non-vanishing result~(\ref{rhov}) can be understood as a Casimir-type of vacuum energy when the boundary conditions and topology play a crucial r\^ole. The estimate~(\ref{rhov}) is based on our understanding of the ghost's dynamics: it can be analytically computed in  the 2d Schwinger model and hopefully it can be tested in 4d lattice QCD as explained in~\cite{our4d,toy}.  Moreover, one can fix the value of $L$ today by equating the observed $\rho_\Lambda$ to~(\ref{rhov}).  This prediction could be directly tested in future cosmic microwave background (CMB) data~\cite{cmbt}.

This contribution to the vacuum energy is computed using QFT techniques in a non-expanding universe without gravity.  As it stands, it cannot be used for studying the evolution of this vacuum energy density with the expansion of the universe. In order to do so one needs to know the dynamics of the ghost field coupled to gravity on a finite manifold.  In this paper we deal with the dynamics of the ghost field when eq.~(\ref{rhov}) is recognised as the strength of its potential energy. We will derive the dynamical equations which govern the evolution of the Veneziano ghost~\cite{veneziano} in a curved spacetime.  The analysis of these equations gives us a time-dependent equation of state $w(t)$ which can be compared with observations.

It is also interesting to note that a somewhat similar estimate was given in 1967 by Zeldovich~\cite{Zeldovich:1967gd} who argued that  $\rho_\Lambda \simeq {m_p^6}/{M_P^2}$: this is numerically of the same order of magnitude as~(\ref{rhov}) if one replaces $L^{-1}\rightarrow H$.  Since then, the form $H\Lqcd^3$ has risen its head several times, see for instance~\cite{Bjorken:2001pe,Schutzhold:2002pr,Bjorken:2004an,Klinkhamer:2007pe,Klinkhamer:2008ns,Klinkhamer:2009nn}.  Despite the apparent similarity between the form $H\Lqcd^3$ and eq.~(\ref{rhov}), it is important to notice that, first of all, in our case it is the \emph{inverse linear size of the embedding compact manifold} which actually appears in eq.~(\ref{rhov}) below, not the Hubble parameter, as in all the aforementioned papers.  This is because our computations are done within QFT on a compact manifold size $L$, while we use $H\sim L^{-1}$ in our numerical estimates. Therefore, our result is of fundamentally different origin.  Finally, let us stress that equation~(\ref{rhov}) is not an estimate based on (although very well motivated) physical arguments, but a result of a precise calculation, which can be explicitly worked out completely analytically in a simplified 2d model defined on a finite manifold such as a torus~\cite{toy}.

The paper is organised as follows. In the next section (section~\ref{ghostlagrangian}) we arrive at the equations of motion which dictate the dynamics of the ghost field.  In section~\ref{interpretation} we discuss the interpretation of the obtained results using the standard lore of QFT which complements the analysis based on classical equations presented in section~\ref{equationofstate}. In particular we argue that the physical Hilbert subspace changes when QCD is coupled to gravity. More specifically, we discuss how the Veneziano ghost being unphysical in the QFT formulation in Minkowski space exhibits non-trivial physics in a non-static universe. The corresponding effects are proportional to the rate of expansion $H$ such that numerically they are naturally very small, $H/ \Lqcd \sim 10^{-41}$.  However, we claim that this ``small'' effect is the driving force which accelerates the universe now.  In other words, acceleration from a vacuum energy with time-dependent equation of state $w(t)$ is a direct consequence of the dynamics of the Veneziano ghost.  Then, in section~\ref{equationofstate} we explicitly obtain the equation of state $w(t)$ by solving the classical equations of motion for our fields.  The classical solutions are analysed (mostly by numerical means) in section~\ref{numerics}.  In the same section we discuss observations and how they relate to our predictions.  Section~\ref{tuning} provides some intuition about the scales involved in the mechanism, and their apparent ``coincidence'' (related to the eponymous problem in cosmology).  We summarise our arguments in the concluding section~\ref{conclude}.

\section{The ghost in Minkowski space: Lagrangian, gauge fixing, quantisation}\label{ghostlagrangian}

The mechanism that bridges the well known Veneziano ghost in QCD to the non-vanishing vacuum energy $\rho_\Lambda\neq 0$ was recently put forward in~\cite{our4d}, and aims at explaining the late accelerating phase of the universe as a result of a vacuum energy term which, in this scheme, is defined as the mismatch between flat and infinite Minkowski spacetime and a topologically non-trivial one.

If the cosmological constant is indeed a result of the spacetime we live in having a non-trivial topology, then there must be a messenger capable of carrying the information about the boundaries (and the related boundary conditions imposed on the quantum fields), and this is possible only when there are strictly massless degrees of freedom which can propagate at very large distances ${\cal O} (H^{-1})$.  We should emphasise that these are not necessary asymptotic physical states (in fact, they are not).  Rather, they should be treated as a convenient way to discuss large distance dynamics, non-trivial boundary conditions, invariance under large gauge transformations etc.  Degrees of freedom of this kind are a handy way to track the physics of the different constraints that arise when the introduction of a auxiliary fields remove some non-localities (anomalies) which always accompany gauge theories.

The crucial for the present work observation is that while na\"ively all QCD degrees of freedom can reach only very short lengths ${\cal O} (\Lqcd^{-1})$, there is a unique (unphysical) degree of freedom which is exactly massless and can instead propagate to arbitrarily large distances: this is the aforementioned Veneziano ghost~\cite{veneziano}.  In short, the Veneziano ghost (see also~\cite{witten}) which solves the $U(1)_A$ problem in QCD is also responsible for a slight difference in energy density between a finite manifold of size $L$ and Minkowski $\mathbb{R}^4$ space, such that $\rho_{\Lambda}\simeq H\Lqcd^3 \sim (10^{-3} {\text eV})^4$.  Notice that this field is not in the physical spectrum, and as such it does not give rise to any of the usual problems associated with negative sign kinetic terms, propagators, norms and commutation relations.

Thus, the correction to the vacuum energy density due to a very large (but not infinite) manifold is small, and goes as $1/L$.  The central point is that although small, this is not exponentially suppressed as $\exp(-L)$, as one could anticipate for any QFT where all physical degrees of freedom are massive (such as in QCD).  Hence, the QCD ghost acts as a source for the cosmological constant, $\rho_{\Lambda}$.  This very small number $(L \Lqcd)^{-1} \simeq H/\Lqcd \sim 10^{-41}$ nevertheless provides a non-vanishing cosmological constant  which is surprisingly close to the observed value as eq.~(\ref{rhov}) explicitly shows.

In this section we will write down the Lagrangian which contains all the dynamical information about the ghost field, and discuss how this field can be integrated out (as was done in the original paper~\cite{vendiv}) to generate the $\eta'$ mass, which was the main result of~\cite{vendiv}.  If we were interested in $\eta'$ physics in Minkowski space, this would be the end of the story.  However, our target is different: we want to obtain these results in Minkowski space keeping an explicit ghost field in the system in order to generalise the construction to a curved spacetime.  It turns out that the Lagrangian put together this way includes another paired massless companion along with the ghost field such that in 4d it \emph{identically coincides}  with the corresponding expression for the 2d Schwinger model~\cite{KS}.  Unitarity in this new description is not violated (and so are not all the usual founding properties of a sensible QFT) thanks to the appearance of the ghost's companion which exactly cancels all unphysical contributions in the same way as it does in the 2d Schwinger model~\cite{KS}. 

Let us remark here that there will be no new physical results when the well known resolution of the $U(1)_A$ problem is simply reformulated in a different way.  However, such a new treatment will be essential and incredibly handy in putting the system into a curved spacetime.  In this case, as we shall see in the next sections, there will be a novel contribution to the vacuum energy arising from the ghost and its partner, as the exact cancellation between these two does not hold anymore.  The entity of this failed cancellation will be proportional to the departure from Minkowski spacetime, i.e. of order $H$.

\subsection{The Lagrangian}

The starting point for our analysis will be the Lagrangian which includes the ghost pole as proposed first by Di~Vecchia and Veneziano in~\cite{vendiv} (a variation on this Lagrangian was already available in~\cite{sch}--we thank J.~Schechter for drawing this important reference to our attention), whose general form reads
\be\label{lag}
{\cal L} = {\cal L}_0&+&\frac{1}{2}\partial_\mu \eta' \partial^\mu \eta'  + \frac{N_c}{b f_{\eta'}^2}q^2 - \left(\theta-\frac{\eta'}{f_{\eta'}} \right)q \\ \nonumber 
&+& N_f m_q |\<\bar{q}q\>| \cos\left[ \frac{\eta'}{f_{\eta'}} \right]  + g.f. \, ,
\ee
where we explicitly keep only the degrees of freedom we are going to be working with, such as $\eta'$ and the topological density $q$, while all others (including $\pi, K, \eta$) are assumed to be in ${\cal L}_0$, and shall not be mentioned in this paper.  In the rest of this section, unless explicitly stated, all quantities are defined in Minkowski spacetime.  In this Lagrangian $g.f.$ means gauge fixing term for three-form $A_{\mu\nu\rho}$, see below, and the coefficient $b\sim m_{\eta'}^2$ is fixed by the Witten-Veneziano relation for the topological susceptibility in pure gluodynamics.  The topological density is defined as usual, 
\be\label{qdef}
q = \frac{g^2}{64\pi^2} \epsilon_{\mu\nu\rho\sigma} G^{a\mu\nu} G^{a\rho\sigma} \equiv \frac{1}{4} \epsilon_{\mu\nu\rho\sigma} G^{\mu\nu\rho\sigma} \, ,
\ee
with
\be
G_{\mu\nu\rho\sigma} &\equiv& \partial_\mu A_{\nu\rho\sigma} - \partial_\sigma A_{\mu\nu\rho} + \partial_\rho A_{\sigma\mu\nu} - \partial_\nu A_{\rho\sigma\mu} \, , \label{four} \\
A_{\nu\rho\sigma} &\equiv& \frac{g^2}{96\pi^2} \left[ A_\nu^a \stackrel{\leftrightarrow}{\partial}_\rho A_\sigma^a - A_\rho^a \stackrel{\leftrightarrow}{\partial}_\nu A_\sigma^a - A_\nu^a \stackrel{\leftrightarrow}{\partial}_\sigma A_\rho^a + 2 gC_{abc} A_\nu^a A_\rho^b A_\sigma^c \right] \, . \label{three}
\ee
The fields $A_\mu^a$ are the usual $N_c^2-1$ gauge potentials for chiral QCD and $C_{abc}$ the $SU(N_c)$ structure constants.  The constant $f_{\eta'}\simeq f_{\pi}$ is the $\eta'$ decay constant ($f_\pi$ being that of the pion), while $m_q$ is the quark mass, and $ \<\bar{q}q\>$ is the chiral condensate.

The three-form $A_{\nu\rho\sigma}$ is an abelian totally antisymmetric gauge field which, under colour gauge transformations with parameter $\Lambda^a$
\be\label{colour}
\delta A_\mu^a = \partial_\mu \Lambda^a + i g C_{abc} \Lambda^b A_\mu^c \, ,
\ee
behaves as
\be\label{transf}
A_{\nu\rho\sigma} &\rar& A_{\nu\rho\sigma} + \partial_\nu \Lambda_{\rho\sigma} - \partial_\rho \Lambda_{\nu\sigma} - \partial_\sigma \Lambda_{\rho\nu} \, , \\
\Lambda_{\rho\sigma} &\propto& A_\rho^a \partial_\sigma \Lambda^a - A_\sigma^a \partial_\rho \Lambda^a \, .
\ee
In this way the four-form $G_{\mu\nu\rho\sigma}$ is gauge invariant.  The term proportional to $\theta$ is the usual $\theta$-term of QCD and appears in conjunction with the $\eta'$ field in the correct combination as dictated by the Ward Identities (WI).  The constant $b$ is a \emph{positive} constant which would give the wrong sign for the mass term of the scalar field $q$, the property which motivated the term ``Veneziano ghost''.  However, this positive sign for $b$ is what is required to extract the physical mass for the $\eta'$ meson, $m_{\eta'}^2 \sim b$, see the original reference~\cite{vendiv} for a thorough discussion.

One should emphasise that the gauge fixing term in~(\ref{lag}) has to be not confused with  the standard gauge fixing term for the conventional gluon field $A_\mu^a$, as it is related to the fixing of the gauge for the three-form $A_{\mu\nu\rho}$ describing the Veneziano ghost and carrying no colour index.  One can interpret the field $A_{\mu\nu\rho}$ as a collective mode which is represented by a specific combination of the original gluon fields, which in the infrared leads to a pole in the unphysical subspace.  We know about the existence of this very special degree of freedom and its properties from the resolution of the famous $U(1)_A$ problem: integrating out the $q$ field (as shown below) provides the mass for the $\eta'$ meson.

One more remark about the coefficient $b$ which enters~(\ref{lag}), and which is the principal ingredient in solving the $U(1)_A$ problem. 
Its magnitude is determined by the topological susceptibility in pure gluodynamics (without quarks) as
\be
\label{top}
\int \!\dd^4x \la T\{q(x), q(0)\}\ra_{\theta=0}=\frac{i f_{\pi}^2}{2N_c} b \, .
\ee
Of course $b=0$ to any order in perturbation theory because $q(x)$ is a total divergence $q = \partial_\mu K^\mu $, see below.  However, as we learnt from~\cite{veneziano,witten}, $b\neq 0$ due to the non-perturbative infrared physics; in fact, $m_{\eta'}^2 \sim b$.  Since we believe this approach to be the correct one, we shall adopt it in our work.

\subsection{Integrating out the topological field $q \simeq \epsilon_{\mu\nu\rho\sigma} G^{\mu\nu\rho\sigma}$}

If one is interested only in the spectrum of the meson particles arising from the low-energy effective Lagrangian~(\ref{lag}), it is convenient to integrate out the scalar field $q$ by means of its equations of motion, since there is no kinetic term associated to it.  This is indeed the procedure followed by Di~Vecchia and Veneziano in their original paper~\cite{vendiv}, the outcome of which, as follows from~(\ref{lag}),  is
\be\label{lag0}
{\cal L} = {\cal L}_0 &+& \frac{1}{2}\partial_\mu \eta' \partial^\mu \eta'  -\frac{bf_{\eta'}^2}{4N_c}\left(\theta -\frac{\eta'}{f_{\eta'}}\right)^2\\ \nonumber
&+& N_f m_q |\<\bar{q}q\>| \cos\left[ \frac{\eta'}{f_{\eta'}} \right] \, ,
\ee
where all the dependence on the three-form $A_{\nu\rho\sigma}$ has disappeared.  This formula explicitly shows that $\eta'$ receives a non-vanishing mass in the chiral limit, $m_{\eta'}^2 \simeq b/2N_c \neq 0$ due to the non-zero magnitude of the coefficient $b$, which enters~(\ref{lag}) and~(\ref{top}).  This formula also reproduces the notorious Witten-Veneziano relation for the topological susceptibility in pure gluodynamics if one substitutes $b=2N_c m_{\eta'}^2$ into the expression~(\ref{top}) for the topological susceptibility.

However, since we are aiming at uncovering the details of the dynamics of the ghost field, we want to reproduce all the physical results reviewed above without integrating out the topological three-form $A_{\mu\nu\rho}$ describing the Veneziano ghost.  As we already mentioned earlier, in a curved spacetime the Veneziano ghost couples to the gravity field; therefore, in curved space, we will not be able to get rid of the ghost, and we will have to deal with it explicitly. In order to simplify this problem, in what follows we shall separate (project out) the relevant longitudinal degree of freedom in three-form $A_{\mu\nu\rho}$; this degree of freedom is the one that contributes to the topological susceptibility~(\ref{top}).

\subsection{Gauge fixing: finding the ghost}

With the grand scheme just outlined in mind, we shall now choose the Lorenz-like gauge 
\be
\label{gauge}
(\partial^{\rho}A_{\mu\nu\rho}) \simeq \left( \partial_\mu K_\nu - \partial_\nu K_\mu \right)=0 \, ,
\ee
in which we will carry out our manipulations.  It is the same gauge which was discussed in the original paper~\cite{vendiv}.  In this formula we have defined $K_{\mu}$ as
\be\label{current}
K_\mu \equiv \epsilon_{\mu\nu\rho\sigma} A^{\nu\rho\sigma} \, , ~~~~~ q = \partial_\mu K^\mu \, .
\ee
We choose to work only with the longitudinal part of the $K_{\mu}$ field because only this longitudinal part determines the topological density $q = \partial_\mu K^\mu$, and eventually leads to a  non-vanishing contribution to the topological susceptibility~(\ref{top}).  Therefore, we  write the longitudinal part of $K_\mu $ as
\be\label{Phidef}
K_\mu \equiv \partial_\mu \Phi \, ,
\ee
such that the expression for the topological density takes the form
\be\label{boxdef}
q = \partial_\mu K^\mu = \Box \Phi \, ,
\ee
where $\Phi$ is a new scalar field of mass dimension 2. We notice that the gauge condition~(\ref{gauge}) is automatically satisfied with our definition~(\ref{Phidef}).  Now our Lagrangian~(\ref{lag}) can be expressed in terms of the $\Phi$ field as follows
\be\label{lag1}
{\cal L} &=& {\cal L}_0 + \frac{1}{2}\partial_\mu \eta' \partial^\mu \eta'  +N_f m_q |\<\bar{q}q\>| \cos\left[ \frac{\eta'}{f_{\eta'}} \right] \\ \nonumber
 &+& \frac{1}{2m_{\eta'}^2f_{\eta'}^2} \Phi \Box \Box \Phi + \left(\frac{\eta'}{f_{\eta'}}\right) \Box \Phi  - \theta \Box \Phi \, ,
\ee
where we plugged in the coefficient $b\rar 2N_c m_{\eta'}^2$ as the Witten-Veneziano relation requires.  If we integrate out the $ \Box \Phi $ field we return to the expression~(\ref{lag0}) which describes the physical massive $\eta'$ field alone.

As usual, the presence of 4-th order operator  $\Phi \Box \Box \Phi$ is a signal that the ghost is present in the system and may be quite dangerous. However, we know from the original form~(\ref{lag}) that the system is unitary, well defined etc, in different words, it does not present any problem associated with the ghost. The lesson we are about to learn will be pivotal in promoting the system to a curved background, when we will not succeed in integrating out the ghost due to the coupling to gravity.

It is convenient to define a new field $ \phi_2$ which is a combination of the the original $\eta'$ field and $\Phi$ as
\be\label{phi2def}
\eta' \equiv \left( \phi_2 +\frac{\Phi}{f_{\eta'}} \right) \, ,
\ee
which serves to complete the squares in~(\ref{lag1}) in such a way that one can eliminate the term  $\int \!\dd^4x \eta' \Box \Phi=- \int \!\dd^4x \partial_\mu \Phi \partial^\mu \eta'$.  The Lagrangian now takes the form
\be\label{lag2}
{\cal L} &=& \frac{1}{2} \partial_\mu \phi_2 \partial^\mu \phi_2 +N_f m_q |\<\bar{q}q\>| \cos\left[ \frac{\phi_2}{f_{\eta'}} +\frac{\Phi}{f_{\eta'}^2}\right]
\\ \nonumber
&+& \frac{1}{2m_{\eta'}^2f_{\eta'}^2} \Phi \left[ m_{\eta'}^2\Box +  \Box \Box \right] \Phi \, .
\ee
It is now straightforward to repeat the known steps in coping with the higher derivatives term $\Box^2$ in~(\ref{lag2}), namely, one recognises that this operator hides an extra degree of freedom, and can be explicitly reduced in terms of the associated propagator as
\be\label{boxsq}
&\left[ \Box \Box + m_{\eta'}^2\Box \right] \tilde\triangle_F = - m_{\eta'}^2\delta^4(x) \, , & \nonumber\\
&\tilde\triangle_F = \lim_{\rho\rar0} \left[ \triangle_F \left( m_{\eta'}, x \right) - \triangle_F \left( \rho , x \right) \right] \, , &
\ee
which is the sum of a massive $m_{\eta'}$  scalar, and a massless ghost-like scalar.  This means that the $\Phi$ field corresponds to two degrees of freedom which is almost an obvious statement if one formally writes the inverse operator as follows
\be
\label{inverse}
\frac{m_{\eta'}^2}{ \Box \Box + m_{\eta'}^2\Box  }= \left(\frac{1}{ - \Box - m_{\eta'}^2  } -  \frac{1}{ - \Box  } \right) \, .
\ee
In analogy with the 2d Kogut and Susskind (KS) model~\cite{KS}, we will call  the massive scalar field as $\hat\phi$  while the massless ghost is $\phi_1$. 
The final Lagrangian, explicitly including all the relevant terms, thus becomes
\be\label{lagKS}
{\cal L} &=& \frac{1}{2} \partial_\mu \hat\phi \partial^\mu \hat\phi + \frac{1}{2} \partial_\mu \phi_2 \partial^\mu \phi_2 - \frac{1}{2} \partial_\mu \phi_1 \partial^\mu \phi_1 \\
&-& \frac{1}{2} m_{\eta'}^2 \hat\phi^2 + N_f m_q |\<\bar{q}q\>| \cos\left[ \frac{\hat\phi + \phi_2 - \phi_1}{f_{\eta'}} \right] \nonumber\, ,
\ee
where all fields have now canonical dimension one in four dimensions.

We claim that the Lagrangian~(\ref{lagKS}) is that part of QCD which describes long distance physics in our context.  There are no new fields or coupling constants beyond the standard model: the Lagrangian~(\ref{lagKS}) is part of it. Notice that~(\ref{lagKS}) is exactly \emph{identical} to that proposed by Kogut and Susskind in~\cite{KS} for the 2d Schwinger model, and therefore, our 4d system~(\ref{lagKS}) is obviously unitary, and satisfies all other important properties of QFT similarly to the 2d Schwinger model. 

As we have shown, the Veneziano ghost in QCD is represented by the $\phi_1$ field in~(\ref{lagKS}) and it is always accompanied by its companion, the massless field $\phi_2$.  This is the basic reason why the Veneziano ghost in QCD does not lead to any disastrous  
consequence in the system as we shall discuss below.

\subsection{Unitarity and the ghost}\label{GB}

As in the KS model, we can readily proceed to quantise this theory, and in order to do so we will impose canonical equal-time commutation relations for the fields $\hat\phi$, $\phi_1$, and $\phi_2$ which follow from~(\ref{lagKS}) as
\be\label{comm}
\left[ \hat\phi (t, \vec{x})\, , \, \partial_t \hat\phi (t, \vec{y})\right] &=& i \delta^3 (\vec{x}-\vec{y})\, , \nonumber\\
\left[ \phi_1 (t, \vec{x})\, , \, \partial_t \phi_1 (t, \vec{y})\right] &=&- i \delta^3 (\vec{x}-\vec{y})\, , \\
\left[ \phi_2 (t, \vec{x})\, , \, \partial_t \phi_2 (t, \vec{y})\right] &=& i \delta^3 (\vec{x}-\vec{y})\, , \nonumber
\ee
whence we evince that $\phi_1$ is a massless ghost field, and its propagator will have an opposite  sign in comparison with the conventional fields.

The cosine interaction term includes vertices between the ghost and the other two scalar fields, but it can in fact be shown~\cite{KS} that, once appropriate auxiliary (similar to Gupta-Bleuler~\cite{G,B}) conditions on the physical Hilbert space are imposed, the unphysical degrees of freedom $\phi_1$ and $\phi_2$ drop out of every gauge-invariant matrix element, leaving the theory well defined, i.e., unitary and without negative normed physical states, just as in the Lorentz invariant quantisation of electromagnetism.  Specifically, this is achieved by demanding that the positive frequency part of the free massless combination $(\phi_2 - \phi_1)$ annihilates the physical Hilbert space:
\be\label{gubl}
(\phi_2 - \phi_1)^{(+)} \left|{\cal H}_{\mathrm{phys}}\right> = 0 \, .
\ee
With this additional requirement the quantum theory built on the Lagrangian~(\ref{lagKS}) is well defined in any respect, and the physical sector of the theory  exactly coincides with~(\ref{lag0}) which was obtained by a trivial integrating out procedure~\cite{vendiv}.  Yet, there is one place where the ghost has physical consequences, and that is the mass spectrum which, through the Witten-Veneziano mass formula~\cite{veneziano,witten} relates the mass of the $\eta'$ to the topological susceptibility of the model.  It is precisely the topological susceptibility which enjoys the uncancelled contribution of the ghost $\phi_1$ with its companion $\phi_2$  because the topological density expressed in terms of $\Phi$, see~(\ref{boxdef}), which is a combination of the ghost $\phi_1$ and the physical massive $\eta'$ field $\hat\phi$, but not of the companion $\phi_2$.  It is this property that leads to a non-cancellation of the topological susceptibility, and eventually to the $\eta'$ mass.

To wrap up this subsection in one statement: we have shown that the Di~Vecchia-Veneziano Lagrangian~(\ref{lag}) is nothing else than the KS Lagrangian promoted to four dimensions, eq.~(\ref{lagKS}).  We are not claiming that we have produced a new result in QCD. Rather we describe the well known resolution of the $U(1)_A$ problem in a different way, with the ghost explicitly and safely present in the system.  As we shall see in the next sections, such a novel point of view is of formidable help when we upgrade to curved space.

\section{The ghost in curved space: physical interpretation}\label{interpretation}
 
\exclude{We now want to interpret what we have obtained so far, and answer the question: what is really happening with the Veneziano ghost and its partner  in the expanding universe?  In this section we shall formulate and answer these questions within QFT (note that in our discussions 
of the ghost dynamics in section~\ref{equationofstate} where the ghost field was treated classically). The QFT analysis will finally elucidate the microscopical nature of the Veneziano ghost in an expanding universe, which remains obscure in the classical analysis.}

In the standard QFT set up the Veneziano ghost is an unphysical degree of freedom, and represents a massless pole in the unphysical subspace of the entire Hilbert space.  As is known, notwithstanding its being unphysical, the ghost still has observable consequences for the physical subspace; for instance, it gives a non-vanishing contribution to the topological susceptibility and therefore solves the famous $U(1)_A$ problem in QCD as Witten~\cite{witten} and Veneziano demonstrated~\cite{veneziano} (see the review~\cite{shore} for a generic introduction, and ref.~\cite{our4d} for a discussion of some specific aspects linked to this paper).

We shall argue below that once QCD is coupled to gravity the ``would be'' unphysical ghost  may lead to another observable consequence (in addition to the one mentioned above). Namely, the ghost field may acquire a vacuum expectation value in an expanding universe, such as our FLRW universe.  All effects related to this ``ghost condensation'' are proportional to the rate of expansion $H$ so that numerically they are naturally very small, $H/ \Lqcd \sim 10^{-41}$.  However, this ``small'' effect is precisely the driving force which accelerates the present-time universe.  The effect of this ``ghost condensation'' in many (formal) aspects is very similar to the well known problem of particle emission in an expanding universe, see below, although with very different interpretation of the associated outcomes.

Finally, we will see how collective modes of the ghost which can be excited in the expanding universe carry very small typical momentum $\omega_k \simeq k \simeq H$ whilst higher frequency modes are strongly suppressed.  Precisely this feature of the spectrum allows us to identify the extra energy with observed  dark energy because the typical wavelengths $\lambda_k$ of the excitations are of the order of the Hubble parameter, $\lambda_k\sim 1/k\sim 1/H\sim 10\textrm{~Gyr}$, and they do not clump on scales smaller than this, in contrast with all other types of matter.  In the limit for which $H \rar 0$, which corresponds to the usual Minkowski QFT formulation, all the modes move back into the unphysical subspace, and we return to the well studied case~\cite{veneziano}.

\subsection{The Veneziano ghost in Minkowski space}

We take off from the relevant part of the Lagrangian for the Veneziano ghost and its partner as derived in section~\ref{ghostlagrangian}.  This Lagrangian governs the dynamics of the massless ghost, $\phi_1$, a massless scalar filed $\phi_2$ and a physical massive meson $\hat\phi$.
\be
\label{lagrangian}
{\cal L} &=& \frac{1}{2} \partial^\mu \hat\phi \partial_\mu \hat\phi + \frac{1}{2} \partial^\mu \phi_2 \partial_\mu  \phi_2 - \frac{1}{2} \partial^\mu \phi_1 \partial_\mu \phi_1 - \frac{m_{\hat\phi}^2}{2} \hat\phi^2 \nonumber\\
&+& N_f m_q|\<\bar{q}q\>| {\cal N} \cos\left[ \frac{ \hat\phi + \phi_2 - \phi_1}{f_{\eta'}} \right] \, ,
\ee
where $\cal N$ means normal ordering.

Going to curved space one just needs to covariantise the derivative operators (which in fact, acting on scalars, are going to reduce to the usual partial derivatives).  The partition function is given by
\be
\label{Z}
{\cal Z} &=& N \int [ {\cal D} \hat\phi ] [ {\cal D} \phi_2 ] [ {\cal D} \phi_1 ] \exp \left\{ i {\cal S} \right\} \, ,\\
{\cal S} &=& \int \dd^4x \sqrt{-g} {\cal L}\, ,
\ee
where for Minkowski space one should put $ \sqrt{-g}=1$.  The bosons satisfy the commutation relations~(\ref{comm}) from whose signs we learn that $\phi_1$ is a massless (but interacting) ghost field, and its propagator will carry a negative sign (in Minkowski space) which is consistent with the negative sign for the kinetic term in~(\ref{lagrangian}).

As we have mentioned in section~\ref{ghostlagrangian}, the theory is well defined once we require that the positive frequency part of the free massless combination $(\phi_2 - \phi_1)$ annihilates the physical Hilbert space:
\be\label{gb}
(\phi_2 - \phi_1)^{(+)} \left|{\cal H}_{\mathrm{phys}}\right> = 0 \, .
\ee

This condition has been discussed in great details in the 2d case of the KS ghost~\cite{KS}, but it can be promptly generalised to our 4d Veneziano ghost as every step is a carbon-copy of the 2d model, only with different forms for the 4d and 2d Green's functions.  Of course, this does not endangers any of the arguments presented in~\cite{KS}.  It is important to notice that the KS ghost in 2d and the Veneziano ghost in 4d enter the expression for the topological susceptibility alone, without their companions $\phi_2$.  The reason behind the topological susceptibility receiving a non-vanishing (negative) contribution from the ghost traces back to this property, and it is in agreement with the WI (in fact, the WI demand so). 

Let us remark here that neither Witten in~\cite{witten} nor Veneziano in his~\cite{veneziano} did discuss the ghost's dynamics because such complication can be entirely sidestepped if one considers only the gauge invariant sector of the theory.  In fact, Di~Vecchia and Veneziano in their original paper~\cite{vendiv} integrated out the $F_{\mu\nu\rho\sigma}$ field even without fixing the gauge; see section~\ref{ghostlagrangian} for details on how this procedure would be implemented in our set up.  The resulting effective Lagrangian would be given by eq.~(\ref{lagrangian}) where now $\phi_1=\phi_2=0$ such that only the massive $\hat\phi$ remains in the system.  In this case this field corresponds to the usual $\eta'$.  As our goal here is not to study the mesons' mass spectrum of QCD, but rather to analyse the dynamics of $\phi_1$ and $\phi_2$, whose physics is much more involved in an expanding universe, once again we will instead (safely) neglect the dynamics of the massive physical $\hat\phi$ field by putting $ \hat\phi=0$.

The subsidiary condition~(\ref{gb}) which defines the physical subspace can recast as
\be\label{gb1}
(a_k-b_k) \left|{\cal H}_{\mathrm{phys}}\right> = 0 \, , \;  \< {\cal H}_{\mathrm{phys}}| (a_k^{\dagger}-b_k^{\dagger}) =0 \, ,
\ee
where we expanded $\phi_1$ and $\phi_2$ on a complete orthonormal basis $u_k (t, \vec{x})$ and $v_k(t,\vec{x})$ as
\be
\label{expansion}
\phi_1 (t, \vec{x})&=&\sum_{k}\left[ a_ku_k(t,\vec{x})+a_k^{\dagger}u_k^*(t,\vec{x})\right] \, , \nonumber\\
\phi_2 (t, \vec{x})&=&\sum_{k}\left[ b_kv_k(t,\vec{x})+b_k^{\dagger}v_k^*(t,\vec{x})\right] \, .
\ee

The equal-time commutation relations~(\ref{comm}) are then equivalent to
\be\label{comm2}
\left[b_k, b_{k'}\right]&=&0 \, , \; [b_k^{\dagger}, b_{k'}^{\dagger}]=0 \, , \; [b_k, b_{k'}^{\dagger}]=\delta_{kk'} \, ,
\ee
for the $\phi_2$ field, whereas for the ghost modes they satisfy
\be
\label{comm1}
\left[a_k, a_{k'}\right]&=&0 \, , \; [a_k^{\dagger}, a_{k'}^{\dagger}]=0 \, , \; [a_k, a_{k'}^{\dagger}]=-\delta_{kk'} \, ,
\ee
where again the sign minus appears in these commutation relations. The ground state $|0\>$ is defined as usual
\be
\label{vacuum}
a_k|0\>=0 \, , ~~~ b_k|0\>=0 \, , ~~~ \forall k \, .
\ee

The sign minus in the commutators~(\ref{comm1}) is known to be carrier of disastrous consequences for the theory if $\phi_1$ is not accompanied by another field $\phi_2$ with properties that mirror and neutralise it.  As thoroughly explained in~\cite{KS}, the condition~(\ref{gb}) or, what is the same,~(\ref{gb1}) are similar to the Gupta-Bleuler~\cite{G,B} condition in QED which ensures that, defined in this way, the theory is self-consistent and unitarity (together with other important properties) is not violated due to the appearance of the ghost.

To see this, one can check that the number operator $\mathrm{N}$ for $\phi_1$ and $\phi_2$ takes the form
\be
\label{N}
\mathrm{N}=\sum_k \left(b_k^{\dagger}b_k- a_k^{\dagger}a_k\right) \, ,
\ee
while the Hamiltonian $\mathrm{H}$ reads
\be
\label{H}
\mathrm{H}=\sum_k \omega_k\left(b_k^{\dagger}b_k- a_k^{\dagger}a_k\right) \, .
\ee
With this form for the Hamiltonian it may seem that the term $- a_k^{\dagger}a_k $ with sign minus implies instability as an arbitrary large number of the corresponding particles can carry an arbitrarily large amount of negative energy.  However, it does not take more paper than a back of an envelope to check that the expectation value for any physical state in fact vanishes as a result of the subsidiary condition~(\ref{gb1}):
\be
\label{H=0}
\< {\cal H}_{\mathrm{phys}}| \mathrm{H} |{\cal H}_{\mathrm{phys}}\>=0 \, .
\ee
In different words, all these ``dangerous'' states which can produce arbitrary negative energy do not belong to the physical subspace defined by eq.~(\ref{gb1}).  The same argument applies to the operator $\mathrm{N}$ with identical result
\be
\label{N=0}
\< {\cal H}_{\mathrm{phys}}| \mathrm{N} |{\cal H}_{\mathrm{phys}}\>=0 \, ,
\ee
where we can see explicitly the pairing and cancelling mechanism at work.
 
In the next subsection we shall see how this simple picture drastically changes when QCD is coupled to gravity.  For our purposes it is sufficient to consider the system of the two fields $\phi_1$ and $\phi_2$ (which represent the Veneziano ghost and its partner, respectively) in a curved background, ignoring all other physical massive  fields such as $\eta'$, Nambu-Goldstones $\pi$, $K$, all glueballs, etc.

\subsection{The Veneziano ghost in curved spacetime}\label{curved}

It is well known that there are inherent subtleties and obstacles when we attempt to formulate a QFT on a curved space~\cite{Birrell:1982ix}.  In this case there is not a natural choice for the set of modes that on which the fields are expanded, these sets being closely related to a more or less ``natural'' coordinate system.  Indeed, the Poincar\'e group is no longer a symmetry of the spacetime and, in general, it would be not possible to separate positive frequency modes from negative frequency ones in the entire spacetime, in contrast with what happens in Minkowski space where the vector $\partial/\partial t$ is a constant a Killing vector, orthogonal to the $t=\mathrm{const}$ hypersurface, and the eigenmodes~(\ref{expansion}) are eigenfunctions of this Killing vector.  The Minkowski separation is maintained throughout the whole  space as a consequence of Poincar\'e invariance.

For our specific problem, i.e., the study of the ghost dynamics in a curved background, these considerations imply that there will be no simple formulation of the physical Hilbert subspace.  Such a formulation in Minkowski space relies on the existence of a well defined positive frequency mode, which is in turn possible because the Killing vector $\partial/\partial t$ is defined uniquely, since the Poincar\'e group is a symmetry of the theory.  A direct consequence of this is the fact that there are no privileged coordinates available in this set up and, therefore, there is no natural mode decomposition similar to~(\ref{expansion}).  This means that a transition from a complete orthonormal set of modes to different one (the so-called Bogolubov's transformations) will always mix positive frequency modes (defined with the annihilation operators $a_k$ and $b_k$) with negative frequency ones (associated with the creation operators $a^{\dagger}_k$ and $b^{\dagger}_k$).  As a result of this mixture, the vacuum state defined by a particular choice of the annihilation operators will not be ``empty'' once we switch back to the original basis defined by $a_k$ and $b_k$.  In other words, in curved space one should generally expect some relevant physical effects due to the ghost modes.  In particular, as we shall see, they can give non-zero contribution to the energy~(\ref{H}). 

Such drastic, profound consequences arising in going from Minkowski to curved space should not be a surprise to anyone who is familiar with the problem of cosmological particle creation in a gravitational background, or the problem of photon emission by a neutral body which is accelerating.  Only very few systems of this kind can be studied and solved exactly, see e.g.\ the reviews~\cite{Birrell:1982ix,dewitt}.  However, the generic picture emerging from these analyses is amazingly simple: the transition from one coordinate system to another leads in general to non-vanishing Bogolubov's coefficients which mix positive and negative frequency modes.  Eventually, it signals a physical production of particles stemming from the interaction with the gravitating background. 

The spectrum of the produced particles as well as the rate of production have been discussed in literature in great details.  The most important outcome that is relevant for our work turns out to be the fact that the typical magnitude of the Bogolubov's coefficients is proportional to the rate at which the background is changing (the Hubble parameter $H$ in case of an expanding universe, or the acceleration rate if we are studying photon emission by a neutral body), and to the total extent of this process, e.g., the total amount of expansion.

The characteristic frequencies of the modes gravitaty can excite in this set up are of order of the Hubble parameter $\omega_k\simeq H$, whereas higher frequency ones are exponentially suppressed.  This last result is easy to understand physically, because one expects the strength of the expansion to be able to excite modes for which $\omega_k \lesssim H$, but not to possess enough energy to reach the higher end of the spectrum, that is, high $k$ modes are only excited very inefficiently.
  
Following the standard technique for the computation of particle production in a curved spacetime we consider, along with the expansion~(\ref{expansion}), a second complete set of--barred--modes
\be
\label{expansion2}
\phi_1 (t, \vec{x})&=&\sum_{k}\left[\bar{a}_k\bar{u}_k(t,\vec{x})+\bar{a}_k^{\dagger}\bar{u}_k^*(t,\vec{x})\right] \, ,\\ 
\phi_2 (t, \vec{x})&=&\sum_{k}\left[\bar{b}_k\bar{v}_k(t,\vec{x})+\bar{b}_k^{\dagger}\bar{v}_k^*(t,\vec{x})\right] \, . \nonumber
\ee
The new vacuum state is defined as 
\be
\label{vacuum2}
\bar{a}_k|\bar{0}\>=0 \, , ~~~ \bar{b}_k|\bar{0}\>=0 \, , ~~~ \forall k \, .
\ee

Now, in order to study the new vacuum, we should expand the new modes $\bar{u}_k$ and $\bar{v}_k$ in terms of the old ones.  Following the notation of the textbook~\cite{Birrell:1982ix} we obtain
\be
\label{bogolubov}
\bar{u}_k&=&\sum_l \left(\alpha_{kl}u_l+\beta_{kl}u^*_{l}\right) \, ,\\
\bar{v}_k&=&\sum_l \left(\alpha'_{kl}v_l+\beta'_{kl}v^*_{l}\right)\, .  \nonumber
\ee
These matrices are called Bogolubov's coefficients, and they can be evaluated as
\be
\label{bogolubov2}
\alpha_{kl}&=&\left(\bar{u}_k, u_l\right) \, , ~~     \beta_{kl}=-\left(\bar{u}_k, u^*_l\right) \, , \\
\alpha'_{kl}&=&\left(\bar{v}_k, v_l\right) \, , ~~     \beta'_{kl}=-\left(\bar{v}_k, v^*_l\right) \, , \nonumber
 \ee
where the brackets define the generalisation of the conventional scalar product for a curved space
\be
\label{product}
\left(\psi_1, \psi_2\right)=-i\int_{\Sigma}\psi_1(x)\overleftrightarrow{\partial}_{\mu}\psi_2^*\sqrt{-g_{\Sigma}} ~ d\Sigma^{\mu} \, ,
\ee
where $d\Sigma^{\mu}=n^{\mu}d\Sigma$ with $n^{\mu}$ a future-directed unit vector orthogonal to the spacelike hypersurface $\Sigma$ and $d \Sigma$  is the volume element in $\Sigma$.  Any complete set of modes which are orthonormal in the product~(\ref{product}) satisfies
\be
\label{orthonormal}
\left(u_k, u_l\right)=\delta_{kl} \, , ~\left(u^*_k, u^*_l\right)=-\delta_{kl} \, , ~\left(u_k, u^*_l\right)=0 \, ,  \\
\left(v_k, v_l\right)=\delta_{kl} \, , ~\left(v^*_k, v^*_l\right)=-\delta_{kl} \, , ~\left(v_k, v^*_l\right)=0 \, .  \nonumber
\ee
Similar relations, of course, are also valid for the $\bar{u}_k$ and $\bar{v}_k$ modes which appear in the alternative expansion~(\ref{expansion2}).  Equating the two expansions~(\ref{expansion}) and~(\ref{expansion2}) and making use of the orthonormality of the modes~(\ref{orthonormal}), one obtains for the annihilation operators
\be
\label{a}
{a}_k&=&\sum_l \left(\alpha_{lk}\bar{a}_l+\beta_{lk}^*\bar{a}^{\dagger}_{l}\right) \, , \\
{b}_k&=&\sum_l \left(\alpha'_{lk}\bar{b}_l+\beta_{lk}^{'*}\bar{b}^{\dagger}_{l}\right) \, . \nonumber
\ee

The Bogolubov's coefficients possess the set of properties
\be
\label{bogolubov-ortho}
\sum_l \left(\alpha_{lk}\alpha^*_{mk} -\beta_{lk}\beta^*_{mk}\right)&=&\delta_{lm} \, , \\
\sum_l \left(\alpha_{lk}\beta_{mk} -\beta_{lk}\alpha_{mk}\right)&=&0 \, , \nonumber \\
\sum_l \left(\alpha'_{lk}\alpha^{'*}_{mk} -\beta'_{lk}\beta^{'*}_{mk}\right)&=&\delta_{lm} \, , \nonumber \\
\sum_l \left(\alpha'_{lk}\beta'_{mk} -\beta'_{lk}\alpha'_{mk}\right)&=&0 \, . \nonumber
\ee
As one can immediately see from~(\ref{a}), the two Hilbert subspaces based on two possible choices of modes $u_k$ and $v_k$, which appear in~(\ref{expansion}), and $\bar{u}_k$ and $\bar{v}_k$, which instead enter in~(\ref{expansion2}), are different as long as $\beta_{kl}\neq 0, \beta'_{kl}\neq 0$.  In particular, the expectation value of the Hamiltonian~(\ref{H}) of the $k$-th state in the barred vacuum $ \<\bar{0}| \mathrm{H}_k | \bar{0}\> $ is
\be
\label{H2} 
\<\bar{0}|  \omega_k\left(b_k^{\dagger}b_k- a_k^{\dagger}a_k\right)| \bar{0}\>= \omega_k\sum_{l}(|\beta_{kl}|^2+|\beta'_{kl}|^2) \, ,
\ee
which is in sharp contrast with eq.~(\ref{H=0}), derived in Minkowski space.

A remark is in order here.  While $a_k^{\dagger}a_k$ partakes in the expression for the Hamiltonian with sign minus, it nevertheless gives a positive sign contribution to the expectation value as a result of an additional minus sign in the commutation relation for the ghost field~(\ref{comm1}).  Hence, no cancellation between the ghost $\phi_1$ and its partner $\phi_2$ could occur in the expectation value~(\ref{H2}), in net contrast with eq.~(\ref{H=0}).

The best way to gain some insight on the result~(\ref{H2}) is to assume that in the remote past and future the spacetime is Minkowskian.  In fact, there is a simple 2d model with a specific profile for the expansion function $a(t)$ interpolating between two Minkowski spacetimes which can be solved exactly.  The outcome (see sec.~3.4 in~\cite{Birrell:1982ix}) is that, even in this plain example $\beta_{kl}\neq0$, which can be understood as a production of particles by the expanding background.  In our case this should not be interpreted as actual emission of ghost modes, as the ghost modes are \emph{not} asymptotic states in Minkowski spacetime, and therefore they can not propagate to infinity; this constrasts with the conventional analysis~\cite{Birrell:1982ix}.  More appropriately, one should interpret~(\ref{H2}) as an additional time-dependent contribution to the vacuum energy in time-dependent background in comparison with Minkowski spacetime.  This extra energy is entirely ascribable to the presence of the unphysical degrees of freedom.  As we mentioned earlier, this is not the first time that a conventionally unphysical ghost
contributes to a physically observable quantity.  A clarifying simple 2d example on this matter can be found in~\cite{eric}.

In more complicated cases when in the remote past and future the spacetime is not Minkowskian, the formulation becomes very subtle as even a natural definition of particles is typically not available (the particle number will not be constant, a fact which makes its measurement inherently uncertain).  We shall not discuss all these subtleties for the fields treated in the present work as they have exactly the same nature as for conventional particles, and they have been extensively discussed in the literature, see~\cite{Birrell:1982ix} and references to the original works therein.  Let us only point out that, while the number of produced particles may be a deceiving concept, and depends on whole prehistory of the spacetime for its interpretation to be sensible, the stress tensor is free from such kind of uncertainties, which is precisely the reason why we computed $\<\bar{0}| \mathrm{H}_k | \bar{0}\>$, see eq.~(\ref{H2}), rather than $\<\bar{0}| \mathrm{N}_k | \bar{0}\>$; see also~\cite{eric}.

In order to understand the general pattern for the behaviour of the Bogolubov's coefficients $\beta_{kl}$ in the expanding universe, it is useful to assume that the expansion is an adiabatically slow process.  In this case the problem can be analysed using a WKB type of approximation~\cite{Birrell:1982ix} similar to the adiabatic approximation in quantum mechanics.  In the context of quantum mechanics it is well known that if the rate of adiabatic expansion is very small (of order $H$), then all the transition matrix elements are proportional to the same rate $H$.  Moreover, the transitions between states with very large difference in energy $\Delta E$, are exponentially suppressed as $\exp(-\Delta E/H)$, see e.g.~\cite{Landau}.

In the case of QFT in a curved spacetime the technique which is required for such an analysis is much more involved than in quantum mechanics, but the basic reason for this suppression has the same nature.  In fact one can argue that the means by which one studies the cosmological production of particles using the Bogolubov's coefficients as discussed above, are similar to instantaneous perturbation theory in quantum mechanics.  Instead, the analysis of the same physical problem using adiabatic vacua is akin to employing adiabatic perturbation theory to calculate transition amplitudes in quantum mechanics.  In any case, the aspect that is essential for the present presentation is the realisation that in all cases the high frequency modes will be strongly suppressed as $\exp(-\omega_k/H)$ in comparison with the expansion rate $H$.

So, the basic result of this section can be formulated as follows: when QCD is coupled to gravity the ``would be'' unphysical ghost, although it still is not an asymptotic degree of freedom, nevertheless contributes to the vacuum energy density in an expanding universe.  This time-dependent ``ghost condensation'' can, for the most part, be regarded as particle emission in an expanding universe, bearing in mind the subtleties reviewed in this section, i.e., no actual particles are being produced).  All such effects are proportional to the rate of expansion $H$, hence, very small, as $H/ \Lqcd \sim 10^{-41}$, and are seen to be related to the tiny momentum $\omega_k \simeq k \simeq H$ available for efficient gravitational interactions, higher frequencies being exponentially suppressed.  We have also seen that when we flatten the spacetime back to Minkowski, all these excited collective ghost's modes will be seized back in the unphysical Hilbert subspace.

The chief ingredient in this discussion is the subsidiary condition~(\ref{gb}) or~(\ref{gb1}), which ensures a peaceful coexistence of widely different scales: $\Lqcd\sim 100$ MeV and $H\sim 10^{-33}$ eV.  We shall elaborate further in section~\ref{tuning} on how a number of fine-tuning issues which most typically plague dark energy models~\cite{ed,turner}, possess simple explanations within the framework advocated in this paper.

\subsection{Ghost dynamics vs.\ Witten's approach}\label{wits}

The ghost we have been working with in this work, and whose activity is central our discussion, was postulated by Veneziano in the context of the $U(1)$ problem.  It is instructive at this point to look at the different approach to that issue as formulated by Witten~\cite{witten}, where the ghost field does not ever enter the system.  

As long as we work in Minkowski spacetime the two constructions are perfectly equivalent as the subsidiary condition~(\ref{gb}) or~(\ref{gb1}) ensures that the ghost degrees of freedom are decoupled from the physical Hilbert subspace, leaving both schemes with the identical physical spectrum.  In a curved space, on the other hand, we argued that the ``would be'' unphysical ghost can produce a positive physical contribution to the energy-momentum tensor.  The question arises naturally: where is the corresponding physics hidden in the language of Witten?

To answer the question we start by recalling the crucial elements of~\cite{witten}.  The main object under focus is the topological susceptibility in pure gluodynamics defined as
\be\label{Q}
\chi\!&\!\equiv\!&\! i\int \!\!\!\dd x   \la 0|T\{Q(x), Q(0)\} |0\ra \, , ~~ Q\equiv \atop \, , \\ \nonumber
Q\!&\!\equiv\!&\!\partial_{\mu}K^{\mu} \, ,~K^{\mu}\!\equiv\frac{g^2}{16\pi^2}\epsilon^{\mu\nu\lambda\sigma}A_{\nu}^a
(\partial_{\lambda}A_{\sigma}^a+\frac{g}{3}f^{abc}A_{\lambda}^bA_{\sigma}^c) \, ,
\ee
We observe that $\chi$ does not vanish in spite of the fact that the operator $Q$ is the total derivative, and therefore $\chi \equiv 0$ in any order in perturbation theory.  The next important element is the sign of $\chi$ which is negative (i.e., it is opposite to what one should expect for a contribution from a conventional physical degree of freedom).  This is precisely the reason for the introduction of the Veneziano ghost which would saturate the topological susceptibility with the required sign.

On the other hand, Witten obtains the same contribution from the contact term which is represented by the equal-time commutator term~\cite{witten}
\be\label{Q1}
\chi_{contact}\!&\!\equiv\!&\! i\int \!\!\! d^3 x   \la 0| \left[K^0 (\vec{x}), Q(0)\right] |0\ra  
\\ \nonumber &=& -\la 0| {\frac{ \alpha_{s}}{4 \pi} G_{\mu \nu}^{a}(0){G}^{\mu \nu a} (0)}| 0\ra
\ee
(see also~\cite{Diakonov} for detailed computations).  The point is that equation~(\ref{Q}) along with the contact term~(\ref{Q1}) determine the dependence of the non-perturbative vacuum energy $E_{vac}$ on $\theta$ as follows from the definition $\chi =- \frac{d^2 E_{vac}}{d\theta^2}|_{\theta=0}$.  Therefore, a non-vanishing value for the topological susceptibility $\chi\neq 0$ with a negative sign solves the $U(1)_A$ problem~\cite{veneziano,witten,vendiv}, (see also eq.~(3) in~\cite{our4d} where the $U(1)_A$ problem is reviewed in this context).

The next step is to consider the system in a curved background.  It has been discussed in length in this paper how this generalisation can be  performed using the ghost degrees of freedom.  If we want to follow Witten's approach instead, the corresponding information does not go away, but rather it is hidden in equation~(\ref{Q1}) where now the correlation function must be considered in the more general background.  Such correlation in a general curved space is not known, nor it is for our FLRW universe (it is the renormalised two-point function of a non-abelian, strongly interacting theory).

In simple geometries such as de Sitter spacetime a few special examples are known, see~\cite{Birrell:1982ix} for an overview,~\cite{Dowker:1975tf} for the free scalar case, and~\cite{allen,woodard} for the abelian vector field case.  In particular, a computation of the simplest expectation value of the scalar field  $\la \phi(0)\phi(0)\ra$ at coinciding points in de Sitter background turns out to be a very complicated problem~\cite{Dowker:1975tf} which requires a delicate subtraction procedure on the Green's function in the given curved background.  The resulting expectation value depends, of course, on the properties of the background field (gravity in this case) such as the rate of expansion $H$.  These are the effects which have been treated in this paper using the Veneziano ghosts.  Conversely, if we were dealing with the solution proposed by Witten, such physics would be shadowed by the properties of the Green's function~(\ref{Q1}) in a curved and time-dependent background and in the associated subtraction prescription.  In Minkowski space this is simply a subtraction constant (as it was introduced by Witten~\cite{witten}), which is just a number.  Generalising to a curved time-dependent background this would be a ``subtraction function'' which, in principle, should be computable from~(\ref{Q1}) after a proper generalisation for the new spacetime.

To conclude this subsection: we do not aim at solving this intriguing but prohibitive task in this work.  Instead, we want to point out that the physical phenomena (described by the Veneziano ghost in our framework) do not disappear when we use a different approach as that advocated by Witten.   The corresponding physics is hidden in the properties of equal-time commutators, which, we suspect, is much more difficult to study in a curved and time-dependent background.

\subsection{Detour to QED}\label{det}

The main goal of this subsection is to demonstrate that a very similar effect related to the ghost degrees of freedom in a time-dependent background is not a specific feature of the Veneziano ghost in QCD, but is to be expected as a generic feature of any gauge theory when some unphysical degrees of freedom are present in Minkowski space, in particular in the well studied case of QED\footnote{The conventional non-abelian Fadeev-Popov ghosts are much more complicated objects and shall not be discussed here.}.

We start with the standard formulation of the Gupta-Bleuler constraint~\cite{G,B} in conventional QED  in Minkowski space: here the positive frequency part of the free massless combination $\partial_{\mu}A^{(+)\mu}$ annihilates the physical Hilbert space
\be
\label{QED_gb}
\partial_{\mu}A^{(+)\mu}| {\cal H}_{\mathrm{phys}}\ra = 0 \, .
\ee
As is known, this additional requirement ensures that $\la {\cal H}_{\mathrm{phys}} | \partial_{\mu}A^{\mu} |  {\cal H}_{\mathrm{phys}} \ra = 0$.  It is important to notice that the condition~(\ref{QED_gb}) is formulated exclusively for the positive frequency part; the negative frequency part is not mentioned in~(\ref{QED_gb}).  In fact, one can not demand a stronger condition such as $\left( \partial_{\mu}A^{\mu}\right)| {\cal H}_{\mathrm{phys}}\ra = 0 $ as it can not be satisfied even for the ground state: the negative frequency half contains creation operators.

When QED is quantised along these lines, all the problems related to the unphysical polarisations $\epsilon^{(0)}$ and $\epsilon^{(3)}$ are automatically straightened out in a way similar to our computations for the expectation value of the energy~(\ref{H=0})~\footnote{In the terminology of the present paper the ghost $\phi_1$ would correspond to the timelike polarisation $\epsilon^{(0)}$--with negative kinetic term--while its partner  $\phi_2$ would be the longitudinal polarisation $\epsilon^{(3)}$.}.  In different words, all these ``dangerous'' states which can produce arbitrary negative energy do not belong to the physical subspace defined by eq.~(\ref{QED_gb}).  All physical states are such that they contain a mixture of photons with polarisations $\epsilon^{(0)}$ and $\epsilon^{(3)}$ such that~(\ref{QED_gb}) is maintained in all gauge invariant scattering matrix elements.
 
Let us now generalise this construction to a curved space (or even in the simpler case when the background is time-dependent).  As is known, in this case there is no natural choice for the set of modes on which the fields are expanded.  In general, it would be not possible to separate positive frequency modes $\partial_{\mu}A^{(+)\mu}$ and write a well defined prescription as in~(\ref{QED_gb}), at least not in the entire spacetime, just as in the Veneziano ghost example.  This is a crucial difference in comparison with Minkowski spacetime when the vector $\partial/\partial t$ is a Killing vector, orthogonal to the $t=\mathrm{const}$ hypersurface, and the eigenmodes in the expansion of the field $A^{\mu}$ are eigenfunctions of this Killing vector.  The separation of positive and negative frequencies modes is maintained throughout the whole space as a consequence of Poincar\'e invariance, and therefore, the Gupta-Bleuler condition (\ref{QED_gb}) is a well defined constraint in Minkowski space in contrast with a time-dependent background when the condition (\ref{QED_gb}) can not be maintained in the entire space.  This means that a transition from a complete orthonormal set of modes to different one will generally mix different frequency modes, and as a result of this mixture, the vacuum state defined by a particular choice of the annihilation operators will not be ``empty'' once we switch back to the original basis.

As for the Veneziano ghost, since a natural definition of particles in a curved background is typically not available (the particle number will not be constant), see~\cite{Birrell:1982ix}, the interpretation in terms of particles is ambiguous.  On the other hand, the energy-momentum tensor is a well defined operator.  It is our interpretation that in the stress tensor we obtain new contributions which are attributable to ``would be'' unphysical polarisations such as $\epsilon^{(0)}$ and $\epsilon^{(3)}$, rather than claiming that these are actual (asymptotic) particles.  Let us stress in passing that both $\epsilon^{(0)}$ and $\epsilon^{(3)}$ contribute with positive sign to expectation value of the Hamiltonian, similar to eq. (\ref{H2}).

As formulated above, this is not a statement about a possible violation of gauge invariance.  Instead, it is a statement about the \emph{global} non-cancellation between unphysical polarisations in a time-dependent background.  Moreover, we expect that in the ``physical'' Coulomb gauge $A_0=0$, $\partial_iA^i=0$ (when one keeps only the physical degrees of freedom) the corresponding effects would also emerge.  However, they will be hidden in the properties of the coinciding points Green's function in a curved background, similar to our discussions in the previous subsection~\ref{wits}.  As is known, these correlators are quite non-trivial and very sensitive to the global properties of spacetime, in contrast with their simple behaviour in Minkowski space.  We expect, however, that this new QED-related contribution to the dark energy would be negligible in comparison with what we have computed above for the Veneziano ghost, because QED effects, by dimensional reasons, are proportional to $H^4\sim (\frac{H}{\Lqcd})^3 \rho_{DE} \sim 10^{-120} \rho_{DE}$.

\section{Ghost classical dynamics and equation of state}\label{equationofstate}

In order to do cosmology with our Lagrangian~(\ref{lagKS}) we have to move the system to a more realistic universe which includes gravity.  We want to analyse the dynamics in a curved background by studying the QCD Lagrangian which explicitly includes the Veneziano ghost~(\ref{lagKS}), supplied by the auxiliary condition~(\ref{gubl}) selecting the physical subspace.  As we demonstrated above, in Minkowski space these two elements faithfully reproduce the well known low-energy dynamics of~(\ref{lag0}).  To this end we will first generalise the relevant piece of the QCD Lagrangian, given by~(\ref{lagKS}), to a general curved background, and then try to solve the corresponding classical field equations for the expectation values of the fields.  This is the standard procedure in studying the dynamics of scalar fields in the early universe, and we are making no exception here.

\subsection{Curved spacetime}

It is in fact merely a matter of rewriting the Minkowski Lagrangian~(\ref{lagKS}) all what one needs to do to promote it to a curved space.  Indeed, its simplicity allows the straightforward generalisation
\be\label{lagC}
{\cal L} &=& \frac{1}{2} D_\mu \hat\phi D^\mu \hat\phi + \frac{1}{2} D_\mu \phi_2 D^\mu \phi_2 - \frac{1}{2} D_\mu \phi_1 D^\mu \phi_1 \\
&& - \frac{1}{2} m_{\eta'}^2  \hat\phi^2 + N_f m_q |\<\bar{q}q\>| \cos\left[ \frac{\hat\phi + \phi_2 - \phi_1}{f_{\eta'}} \right] \nonumber\, ,
\ee
where the covariant derivative $D_\mu$ is defined as $D_\mu = \partial_\mu + \Gamma_\mu$ so that, for instance $D_\mu V^\nu = \partial_\mu V^\nu + \Gamma_{\mu\lambda}^\nu V^\lambda$.

One can easily check that all the steps that led to the Minkowski Lagrangian~(\ref{lagKS}) can be just as well worked out for a general background, see~\cite{toy} for an example.  Therefore, ignoring the $\theta$-term contribution, the action for our system will be given by ${\cal S} = \int\!\dd^4x\sqrt{-g}\,{\cal L}$, where $g \equiv \det{g_{\mu\nu}}$ is the determinant of the metric.

Accordingly, the energy-momentum tensor for the system of the $\hat\phi$, $\phi_1$, and $\phi_2$ fields is
\be\label{emtensor}
T_{\mu\nu} &=& - \frac{2}{\sqrt{-g}} \frac{\delta {\cal S}}{\delta g^{\mu\nu}} \\
&=& - D_\mu \hat\phi D_\nu \hat\phi - D_\mu \phi_2 D_\nu \phi_2 + D_\mu \phi_1 D_\nu \phi_1 + g_{\mu\nu} {\cal L} \nonumber \, ,
\ee
which satisfies the energy conservation equation $D^\mu T_\mu^\nu = 0$ as it must.  The equations of motion on which the stress tensor is conserved are cast as
\be
D^\mu D_\mu \hat\phi \!&+&\!\! m_{\eta'}^2 \hat\phi + \frac{N_f m_q |\<\bar{q}q\>|}{ f_{\eta'}} \sin\!\left[ \frac{\hat\phi + \phi_2 - \phi_1}{f_{\eta'}} \right] = 0 \, , \nonumber\\
&&\label{eomhat}\\
D^\mu D_\mu \phi_2 \!\!&+&\!\! \frac{N_f m_q |\<\bar{q}q\>|}{ f_{\eta'}} \sin\!\left[ \frac{\hat\phi + \phi_2 - \phi_1}{f_{\eta'}} \right] = 0 \, , \label{eomphi2}\\
D^\mu D_\mu \phi_1 \!\!&+&\!\! \frac{N_f m_q |\<\bar{q}q\>|}{ f_{\eta'}} \sin\!\left[ \frac{\hat\phi + \phi_2 - \phi_1}{f_{\eta'}} \right] = 0 \, .\label{eomphi1}
\ee
Notice how the combination $(\phi_2 - \phi_1)$ is special in that it is a plane wave (in Minkowski space):
\be\label{plane}
D^\mu D_\mu \left( \phi_2 - \phi_1 \right)=0 \, .
\ee
We suggest that these equations of motion are the interesting part of QCD describing long distance physics in a curved spacetime.  There are no new fields or coupling constants beyond the standard model.

\subsection{Solutions}

Let us now follow in some details the passages which lead to an (approximate) solution of these differential equations.  From now on we will make the simplifying assumptions that the field $\hat\phi$ sits at the bottom of its potential, and we fix its expectation value to zero.  Moreover, we will be treating these fields only from a classical point of view (although we omit the angle brackets), and discuss the evolution of their vacuum expectation values (vevs) in the general background $g_{\mu\nu}$, which for definiteness we choose to be of Friedmann-Lema\^itre-Robertson-Walker (FLRW) type--see however appendix~\ref{Atorus} for a generalisation involving compact 3d hypersurfaces (as required for consistency in our set up)--and hence given by
\be\label{flrw}
\dd s^2 = \dd t^2 - a(t)^2 \left( \dd x^2 + \dd y^2 + \dd z^2 \right) \, ,
\ee
where we denote the scale factor of the universe as $a$.  For future reference, we define the Hubble parameter as $H = \dot a/a$ where the overdot is a derivative with respect to $t$.

Let us change names for the two fields $\phi_1$ (the ghost) and $\phi_2$ (its partner) to $\psi$ and $\varphi$, respectively in order to emphasise that we treat $\psi$ and $\varphi$ as classical fields in this subsection. Their quantum nature has been discussed in section~\ref{interpretation}.  The classical treatment of the system is by far the widest chosen approach in dealing with dark energy matters, and this is the reason why we present this formalism here.  However, while in other, perhaps more familiar, cosmological models such as inflation, the passage from quantum to classical is justified \emph{a posteriori} (see for instance the discussion in~\cite{liddle}, section 7.4.7, and reference to original works therein), and is a fundamental ingredient for the success of the inflationary theory, in coping with our quantum fields we do not expect such a ``little miracle'' to happen.  In other words, the quantum nature of our fields which appears in their non-trivial dependence on the gravitational background as well as on the global properties of the manifold, is brought in at the level of the renormalisation procedure, and is therefore not describable in a purely classical approach.  Therefore, all results based on the classical equation of motion presented in the next two sections should be taken very cautiously, as we do not have any theoretical arguments supporting the consistency of the classical approach (in contrast with the case of inflation).  Still, we will make an attempt in this direction, these inherent difficulties notwithstanding, and address in more details the reasons for our mistrust in this na\"ive construct thereafter.

We want to work in conformal time $\tau$ for the moment, which amounts to the coordinate transformation $\dd \tau = \dd t / a$.  Finally, we will initially solve for the rescaled fields $\tilde\varphi$ and $\tilde\psi$ defined as
\be\label{resdef}
\tilde\varphi \equiv a \varphi \quad,\quad \tilde\psi \equiv a \psi \, .
\ee
Primes will mean derivatives with respect to conformal time $\partial/\partial\tau$.

With these definitions one can either recast the action or act directly on the equations of motion.  In any case, the explicit equations of motion for the tilded fields are 
\be
\tilde\varphi'' - \frac{a''}{a} \tilde\varphi + \omega^2 \left( \tilde\varphi - \tilde\psi \right) &=& 0 \, , \label{eomphit}\\
\tilde\psi'' - \frac{a''}{a} \tilde\psi + \omega^2 \left( \tilde\varphi - \tilde\psi \right) &=& 0 \, , \label{eompsit}
\ee
where to simplify things we keep only the linear term in the expansion, $\sin (\phi_2-\phi_1) \simeq (\phi_2-\phi_1)$.  This approximation is more than sufficient for our qualitative discussions. The  $a$-dependent mass-like term is defined as
\be\label{omega}
\omega(a)^2 \equiv \frac{N_f m_q |\<\bar{q}q\>|}{(m_{\eta'} L) (f_{\eta'}a)^2} a^4 \, .
\ee

A few comments on this potential are in order here.  First of all, the $a''/a$ term is nothing else than the usual friction term in the original time $t$ coordinatisation, for the original fields $\varphi$ and $\psi$.  This becomes an effective time-dependent mass term when one employs conformal coordinates and rescales the fields as in~(\ref{resdef}), and is proportional to the Ricci scalar curvature $R$.  In our Lagrangian the fields have been chosen to be minimally coupled to gravity, but in general this coupling will appear and be of the form $(\xi_\varphi \varphi^2 R - \xi_\psi \psi^2 R)$, and in the case of conformal coupling will make the $a''/a$ term in~(\ref{omega}) cancel.

Second, the reader will have noticed that there is an extra factor of $1/(m_{\eta'}L)$ in the second term. This is a very subtle point which we have thoroughly discussed in our previous paper~\cite{our4d}, and it is a result of a subtraction procedure that compares the values of the vacuum energy in Minkowski space with that in a general compact manifold (such as a torus of size $L$).  This prescription aims at extracting the physical and measurable portion of the vacuum potential energy of the ghost field, by taking the difference between the vacuum energy in compact space and that in infinite Minkowski space, and is of strict quantum mechanical origin.  Essentially, this is our definition of the vacuum energy when the ``renormalised energy density'' is proportional to the departure from Minkowski spacetime geometry and remains finite.  This external (to the classical system) prescription will necessarily introduces some inconsistency in the classical picture, as we will see below.

Notice that the correction (which was computed exploiting the topological susceptibility of QCD when the ghost is present) is \emph{linear} in the inverse size of the manifold, not exponentially suppressed, as one would generally expect in the confined phase of QCD.  This contribution is a direct result of the existence of the ghost and can be computed in an exact way in a compact 2d spacetime in the context of the Schwinger model~\cite{toy}, and indeed shows the linear dependence we have used in~(\ref{omega}).  There are very strong arguments to believe that this is going to be the case in 4d as well, for the topological structure (with the contribution of the ghost to the topological susceptibility as \emph{demanded} by the WI) is identical in both models~\cite{our4d}, and because as we have just shown in sec.~\ref{ghostlagrangian}, the low-energy Lagrangians coincide~(\ref{lagKS}).  There is therefore no arbitrariness in this choice of the potential: in fact, this is not a choice, as the potential is the one that correctly captures the chiral dynamics of low-energy QCD.

Since ultimately we are interested in the combination $(\dot\varphi^2 - \dot\psi^2)$ which enters the expression for the equation of state, we look for solutions of the corresponding combination in conformal time for the rescaled fields.  Even in curved space, the solution for $(\tilde\varphi - \tilde\psi)$ is straightforward.  Assuming the fields do not depend on the three spatial coordinates we find
\be\label{minust}
\frac{\partial^2}{\partial\tau^2} \left( \tilde\varphi - \tilde\psi \right) = 0 \quad \Rar \quad \left( \tilde\varphi - \tilde\psi \right) = c_0 + c_1 \tau \, ,
\ee
with $c_0$ and $c_1$ the integration constants.  The following step is to use this solution in the equation for $\tilde\varphi$~(\ref{eomphit}) which becomes
\be\label{eomphit2}
\tilde\varphi'' = - \omega^2 \left( c_0 + c_1 \tau \right) \, .
\ee

This equation can be solved analytically if we assume that $R \ll \Lqcd/L$, as is certainly the case in the late-time regime we are interested in, in which case the curvature term in $\omega^2$ can be safely neglected.  The time dependence of $\omega^2$ then simplifies greatly, being the same as that of $a^4 / (a^2 L) \propto a(t)$.  To keep things simple we also choose $c_1 = 0$; of course the results we will present in the following section can only be generalised by retaining the full form for $(\tilde\varphi - \tilde\psi)$.  Equation~(\ref{eomphit2}) is then solvable for the first derivative of the field, and so is its counterpart for the original field $\varphi(t)$, whose solution reads
\be\label{phisol}
\dot\varphi(t) = \frac{\dot\varphi(t_0)}{a^2} - \varphi H + \frac{\varphi(t_0) H_0}{a^2} - \frac{\omega^2(t_0) }{a^2} c_0 (t - t_0) \, ,
\ee
where $t_0$ is the present time, and obviously the conformal length $L$ scales as $L = L_0 (a(t) / a(t_0))$.  We can simplify this expression further, as long as we are working in the late-time regime (that is, much later the QCD phase transition epoch), by noticing that if the characteristic vev of the scalars $\varphi$ and $\psi$ is of order of the QCD scale (which we can parametrise as $\Lqcd$ or with $f_{\eta'}$), although small enough to justify the simplification of the cosine interaction term, then the first of all the terms in equation~(\ref{phisol}) will dominate.  One implicit assumption we are also making is that the physical size of the manifold $L_0$ is of order (but still larger) of the Hubble length, $H_0$, as required by observations, see the discussion in~\cite{cmbt} (if this were not the case, then also the term $\omega^2(t_0) c_0 t_0 / a^2$ would give a similar contribution--notice that the dependence on time, through $a$, is the same).

As mentioned earlier, we are aiming at finding the explicit expression for $(\dot\varphi^2 - \dot\psi^2)$, which leads us to the following expression
\be\label{minus}
\dot\varphi^2 - \dot\psi^2 \!&=&\! \left( \dot\varphi + \dot\psi \right) \, \left( \dot\varphi - \dot\psi \right) \nonumber\\
\!&=&\! \left[ 2\dot\varphi - \frac{c_1}{a^2} + \frac{c_0 + c_1 \tau}{a} H \right] \, \left[ \frac{c_1}{a^2} - \frac{c_0 + c_1 \tau}{a} H \right] \nonumber\\
\!&\simeq&\! - 2 \dot\varphi(t_0) \frac{c_0 H}{a^3} \, ,
\ee
according to the simplifications explained in the previous paragraph.

As a last step, let us recall what $c_0$ is.  In equation~(\ref{minust}) we have defined it to be a constant in conformal time for the rescaled fields $\tilde\varphi$ and $\tilde\psi$.  This means that once we switch back to the physical fields, $c_0$ will not be constant anymore, but is now proportional to $a(t)$, and in our normalisation where the initial conditions are given today, it is going to be proportional to $a_0 \varphi(t_0) \simeq a_0 f_{\eta'}$.  This is going to be the case for the other contribution coming from $\dot\varphi(t_0)$ which is going to be of order $f_{\eta'}^2$.  All together, since we do not know the position of the field in its potential now, nor its velocity, it is simplest to parametrise them by means of one and the same constant (which we therefore expect to be somewhat smaller than 1) baptised $c_\varphi \simeq \varphi(t_0) \dot\varphi(t_0) / f_{\eta'}^3$ (in principle positive or negative).

The final expression for the kinetic term for the two fields $\varphi$ and $\psi$ (or $\phi_2$ and $\phi_1$, which is equivalent) is simplified to
\be\label{kinetic}
\frac{1}{2}\dot\varphi^2 - \frac{1}{2}\dot\psi^2 = \alpha_1 \frac{H}{H_0} \left( \frac{a_0}{a} \right)^3 \, ,
\ee
where we have defined $\alpha_1 \equiv 2 c_\varphi f_{\eta'}^3 H_0$.  A few remarks on the estimate~(\ref{kinetic}) are in order.  First, as expected, this contribution is the same order of magnitude, $\Lqcd^3 H_0$, as the potential term, $\Lqcd^3/L_0$, discussed previously~\cite{our4d}; both contributions are of the same order of magnitude of the critical density $\rho_c = 3H_0^2M_P^2$. Second, as we underlined previously, there are no new fields and coupling constants involved in our analysis, as everything is within QCD.  The unknown coefficient $c_\varphi \lesssim 1$ which appears in our description is not a consequence of unknown physics. Rather, it is determined by the initial conditions when $\varphi(t)$ and $\psi (t)$ are treated classically.

Lastly, the result~(\ref{kinetic}) may look suspicious as it obviously violates the subsidiary condition~(\ref{gubl}). Let us remind that  this condition  was a key element leading to the cancelation of the ghost contribution $\phi_1$ with its companion $\phi_2$ such that a QFT based on~(\ref{lagKS}) is well defined in all respects as explained in section~\ref{GB}.  While the condition~(\ref{gubl}) is indeed violated in a curved space, we should notice that the level of its violation is proportional to the rate of expansion $H$ such that in a Minkowski spacetime, when $H=0$, we again obey~(\ref{gubl}).  Moreover, according to the logic of this paper, the breaking of the requirement~(\ref{gubl}) in curved space, which ultimately is proportional to $\Lqcd^3 H_0$, is part of the energy density that accelerates our universe today. Why and how the physical subspace changes when the system is defined in a curved background has been explained in depth in section~\ref{interpretation}, where we explicitly demonstrate that the vacuum expectation value of the Hamiltonian representing the ghost field (and its companion) in such set up does not vanish.  We shall argue that it is instead proportional to the expansion rate $H$, which is consistent with our estimate~(\ref{kinetic}) based on the analysis of the classical equations of motion.

\subsection{Equation of state}

We have finally come up with an analytic and well defined expression for the \emph{sum} of the kinetic energy densities of the two fields $\varphi$ and $\psi$, eq.~(\ref{kinetic}).  Their total energy density (comprising the potential term as well) is found as usual by inspecting the energy-momentum tensor~(\ref{emtensor}), which for a perfect homogeneous and isotropic fluid can be written in our conventions as
\be\label{pf}
T_{\mu\nu} = - \left( \rho + P \right) u_\mu u_\nu + P g_{\mu\nu} \, ,
\ee
where $\rho$ is the energy density and $P$ the pressure density of some component; $u_\mu$ is the proper 4-velocity of the fluid.  There are several subtleties related to the use of this perfect fluid form in the context of this work: they are unfolded partly in section~\ref{numerics}, and more thoroughly in appendix~\ref{Atorus}; we shall employ this simple form for now.

The total (kinetic plus potential) energy density and pressure density of the vacuum are found to be just as usual
\be
\rho_\Lambda &=& \alpha_1 \frac{H}{H_0} \left( \frac{a_0}{a} \right)^3 + \alpha_2 \frac{a_0}{a} \, , \label{rho}\\
P_\Lambda &=& \alpha_1 \frac{H}{H_0} \left( \frac{a_0}{a} \right)^3 - \alpha_2 \frac{a_0}{a} \, , \label{P}
\ee
with the second (still to be determined) constant appearing above being defined as $\alpha_2 \equiv (N_f m_\pi^2 f_\pi^2) / (4 m_{\eta'} L_0)$ (we have rewritten the chiral condensate in terms of measured parameters $m_\pi$ and $ f_\pi$ as $m_q |\<\bar{q}q\>| = m_\pi^2 f_\pi^2 / 4$).

The two expressions~(\ref{rho}) and~(\ref{P}) can be combined in $w = P_\Lambda/\rho_\Lambda$ to yield the effective equation of state $w$ of the system with the Veneziano ghost and its partner (which is identified with the equation of state for dark energy), and which can be written as
\be\label{eos}
w(a) = \frac{\alpha_1 \frac{H}{H_0} \left( \frac{a_0}{a} \right)^3 - \alpha_2 \frac{a_0}{a}}{\alpha_1 \frac{H}{H_0} \left( \frac{a_0}{a} \right)^3 + \alpha_2 \frac{a_0}{a}} = \frac{H/H_0 - a^2 \alpha_2/\alpha_1}{H/H_0 + a^2 \alpha_2/\alpha_1} \, .
\ee
The equation of state~(\ref{eos}) possesses a highly non-trivial time dependence as $H$ in the r.h.s.\ of the equation implicitly depends on $w(a)$. As for the well known scalar field this equation of state oscillates between -1 and 1 as long as the constants $\alpha_1$ and $\alpha_2$ have the same sign (see however the discussion in section~\ref{numerics} and appendix~\ref{Atorus}).  We will be able to find solutions for a negative value of this ratio as well, which would imply an equation of state $w < -1$, which implies a phantom-type of field, and which could be useful in the construction of bouncing cosmologies (see the review~\cite{bounce} with an extensive references list of the original proposals).

We are going to study this equation of state in section~\ref{numerics}, explicitly solving it by numerical means within the redshift range $0 < z < 2$, which is the range spanned by direct measurements of the Hubble parameter, and therefore is the most interesting and constraining for the comparison of the model against the data. 

A final comment before moving on to the numerics of the Veneziano ghost.  As we mentioned earlier, the numerical value of the constant $\alpha_1$ which triggers its late-time magnitude is of the same order of magnitude as the potential part, see~(\ref{potential}) below.  This is why one should be careful in using the estimate~(\ref{rhov}) to directly infer the equation of state of this vacuum energy in an expanding universe, which would be mistakenly deduced to be $w = -2/3$ (corresponding to $a^{-1}$), while it is actually determined by~(\ref{eos}). Furthermore, there is one further subtlety in this respect, which is elaborated on in section~\ref{numerics}.

\section{Numerical analysis: confronting with observations}\label{numerics}

\subsection{Background}

The reference equation in this section is the definition of the time-dependent equation of state~(\ref{eos}).  This equation is highly non-trivial to solve, because the equation of state $w(a)$ appears in the r.h.s.\ as well, being hidden in the explicit expression for $H$.  Moreover, since in general $w$ depends on $a$ as we have seen,  $w=w(a)$, it appears inside the integral which gives the evolution of the vacuum energy density with time.

The Hubble parameter is given by
\be\label{hubbleP}
H^2 = H_0^2 \left[ \Omega_M^0 (a/a_0)^{-3} + \Omega_\Lambda^0 e^{-3 \int \!\dd a (1+w)/a} \right] \, .
\ee
The density fractions $\Omega_M^0$ and $\Omega_\Lambda^0$ are today's values of $\Omega_M \equiv \rho_M / \rho_c$ and $\Omega_\Lambda \equiv \rho_\Lambda / \rho_c$, respectively.  Here $\rho_c \equiv 3 M_P^2 H^2$ is the critical energy density for which the 3d space is flat.  The suffixes $M$ and $\Lambda$ refer to matter (the sum of visible and invisible cold matter, mostly in the form of dark matter) and vacuum, respectively.  The Hubble parameter as given in~(\ref{hubbleP}) comes from the Friedmann equation
\be\label{friedmann}
H^2 = \frac{\rho}{2M_P^2}  - \frac{\kappa}{a^2} \, ,
\ee
with $\kappa$ the 3d curvature, and $\rho$ is the sum of all the energy densities in the universe.  Hence, 3d flat space corresponds to $\kappa = 0$ which implies $\Omega = 1$ where, analogously, $\Omega = \sum_i \Omega_i$ ($i$ indicises the different components of the universe).

For the purpose of the numerical study we are going to present it is particularly convenient to normalise the scale factor to today's value $a_0$, and we will also similarly normalise the Hubble parameter to $H_0$.  In this case one can rewrite the equation of state~(\ref{eos}) as
\be\label{eosN}
w(a) = \frac{\left[ \Omega_M^0 a^{-3} + \Omega_\Lambda^0 e^{-3 \int \!\dd a (1+w)/a} \right]^{1/2} - \alpha a^2}{\left[ \Omega_M^0 a^{-3} + \Omega_\Lambda^0 e^{-3 \int \!\dd a (1+w) /a} \right]^{1/2} + \alpha a^2} \, ,
\ee
where  the constants $\alpha_1$ and $\alpha_2$ and their ratio $\alpha$ are defined as
\be\label{alphas}
\alpha_1 \equiv 2 c_\varphi f_{\eta'}^3 H_0, ~~ \alpha_2 \equiv \frac{N_f m_\pi^2 f_\pi^2}{4 m_{\eta'}} c_g H_0,~~ \alpha \equiv \alpha_2/\alpha_1.
\ee
In the preceding definition for $\alpha_2$ we follow~\cite{cmbt} and compare the unknown size of the manifold $L_0$ to the Hubble parameter today $H_0$ via $L_0 = (c_g H_0)^{-1}$, with $c_g$ an unknown constant of order one.  Let us stress once more that this unknown constant, as was for $c_\varphi$, is not a consequence of unknown interactions of unknown fields.  Rather, $c_g$ is determined by the size of the manifold we live in as explained in~\cite{our4d,cmbt}.  As one can see from~(\ref{eosN}) the equation of state $w(a)$ is only sensitive to the ratio of these two unknowns $c_\phi$ and $c_g$.

\subsection{Subtleties} 

A few comments on equation~(\ref{eosN}) are in order.  As the reader will have noticed, we are using an expression for the potential energy of the two fields $\phi_1$ and $\phi_2$ that comes from a very specific subtraction procedure defined in compact non-expanding spacetime~\cite{our4d}.  There are two subtleties of this prescription which affect  our modelling.\\

{\bf a).} First of all, the subtraction in Minkowski spacetime gives the potential energy an inverse linear dependence on the size of the manifold we started out with.  This is fully analytically trackable in 2d~\cite{toy}, and is certainly present in non-expanding 4d QCD as we argued previously~\cite{our4d}.  However, the system we are dealing with in this paper is embedded in a much more complicated set up, where gravity now partakes in the dynamics.  It is licit to ask what would be the result of the same subtraction prescription when applied to a non-static case.

Ideally, one would write down the most general form for the propagators of our fields $\phi_1$, $\phi_2$ and $\hat\phi$ in the general FLRW background \emph{in the compact space}, with the most general dependence on the parameters that define the basic manifold (known as the Teichm\"uller parameters, see appendix~\ref{Atorus}).  Once the propagators have been computed for compact and infinite FLRW spacetime, the subtraction procedure should be implemented, and the result would be the exact form of the ``physical'' potential energy density as defined in our framework.  Notice that our ``point of renormalisation'' in the end will be Minkowski space, in such a way that the subtraction is performed in two steps, first in one and the same spacetime extracting the physical Casimir energy, and second renormalising to Minkowski space.

Unfortunately this is a stunningly complicated task, and we are not able to tackle it analytically.  However, it is easy to guess which kind of form the exact expression in a compact FLRW spacetime will be, or at least, what its principal contribution would be.  Indeed, there are few dimensional parameters describing our system: the Hubble parameter, the sizes which define the manifold (of which we pick a characteristic size $L_0$), and QCD parameters of order $f_{\eta'}$.  In general, starting from the original expression
\be\label{potential}
V_{\mathrm{ghost}} =  \frac{N_f m_\pi^2 f_\pi^2}{m_{\eta'} L} \, ,
\ee
the corrections due to the expansion would be proportional to some  dimensionless combinations of the aforementioned parameters, which basically means that eq.~(\ref{potential}) will be multiplied by some function of $x\equiv HL\sim 1$ as the two other possible combinations $H/f_{\eta'}$, or $(L f_{\eta'})^{-1}$ are very small, ${\cal O} (10^{-42})$, and can be safely neglected.  Therefore, we phenomenologically introduce a function $f$ with the intent to parametrise the outcome of the ``ideal'' subtraction: $f(x) = f(HL)$. Remember that $H$ and $L$ are both normalised to present day values.

The exact form of this function $f(HL)$ is unknown, and should be fixed by the fully consistent calculation of the propagators for the ghost and its partner in compact FLRW universes as described above.  In this work we shall limit ourselves to impose some physical conditions on this function which would allow us to write down a representative form for it without actually knowing it.  The requirements this function must meet are: it has to reproduce our Minkowski findings when the curvature is zero, that is, $f(x) \rar 1$ when $x \rar 0$; second, when the manifold size $L \rar \infty$ the dependence on the size itself in the energy density must disappear, which means that in the limit $x \rar \infty$ we ask that $f(x) \rar x$.  A function which is compatible with these demands is
\be\label{fx}
f(x) \equiv 1 + x \, , ~~~~ x\equiv HL\, , 
\ee
and the potential is accordingly modified in everywhere it appears (most importantly, in~(\ref{eosN})) as
\be\label{potentialf}
V_{\mathrm{ghost}} =  \frac{N_f m_\pi^2 f_\pi^2}{m_{\eta'} L} f(HL)\, .
\ee
We should point out that the introduction of $f(HL)$ is a phenomenological way to account for the unavoidable mismatch that is general would arise between the equation of state determined by our equations of motion in the FLRW background, and the simple promotion of the Minkowski result $\propto 1/L$ to the new background, having treated the potential by means of our subtraction procedure in a non-expanding universe only.

{\bf b).} There is a second subtle point in the expression~(\ref{eosN}), namely the fact that it has been derived assuming a perfect isotropic fluid's stress tensor, while it has been pointed out by us previously~\cite{our4d,cmbt} that we need to work in a generally non-isotropic compact manifold.  For instance the three-torus $T^3$ can easily be not isotropic, as its physical fundamental sizes can be all different, and the expansion rates be unequal in each direction, see appendix~\ref{Atorus}.  The perfect fluid form may not be justified for use in a sufficiently nontrivial manifold, and this will introduce more parameters in the equation of state, which for simplicity are not discussed here.  Nevertheless, this introduces yet another source of error in our numerical solutions shown below.

{\bf c).} The arguments outlined in the previous section about the validity of the classical analysis apply here, and inevitably lead to inconsistencies in some of the formulas we will be employing below.

All this should be kept in mind in playing with the seemingly innocuous equations describing the system.  The missing information about the manifold is going to appear in the numerical results in forms of inconsistencies between different parametrisations of the same quantities which ordinarily would be equivalent.  As an example, the main discrepancy is going to be found in the expressions for the energy density, which can be directly calculated using the classical solutions of the equations of motion eq.~(\ref{rho}) or by parametrisation through the effective equation of state found in~(\ref{eos}).  These subtleties are the topic of the appendices, and we postpone the details until then.  Dealing with these uncertainties, and minimising their impact results in fixing the only two free parameters in this formulation, that is $c_\phi$ and $c_g$: the precise way this is done is explained in appendix~\ref{Anum}, and here we only employ the outcome of that analysis, that is
\be\label{const}
c_\phi \approx 1.1 \cdot 10^{-3} \quad \mathrm{and} \quad c_g \approx 5.1 \cdot 10^{-2} \, .
\ee
Recall that $c_\phi$ can be interpreted as the position of the ghost field in its potential times its velocity at the present time.  This interpretation, and the numbers given above, are consistent with the approximations we have made in deriving the solutions of the equations of motion.  Finally, $c_g$ parametrises the size of the manifold in terms of the Hubble parameter today, which, for~(\ref{const}) means that $L \simeq 19 H_0^{-1}$, and consequently safely beyond the last scattering surface~\cite{cmbt}.

\subsection{Results}\label{results}

We plot in figure~\ref{eosPlot} the equation of state $w(a)$ as a function of the scale factor of the universe, when the $c$ constants are fixed as in~(\ref{const}), for the redshift range $0\lesssim z \lesssim 2$, where the redshift is defined as $1+ z \equiv 1/a$ (remember that we normalise everything today).  The numbers we adopt in our numerical analysis, beside~(\ref{const}), are:
\be\label{somenumbers}
\begin{array}{ll}
m_\pi = 135 \, \mathrm{MeV} \, , &H_0 = 2.13 h \cdot 10^{-42} \mathrm{GeV} \, , \\
m_{\eta'} = 958 \, \mathrm{MeV} \, , & h = 0.71 \, , \\
 f_\pi = 133 \, \mathrm{MeV} \, , & \Omega_\Lambda = 0.73 \, , \\
 f_{\eta'} = 104 \, \mathrm{MeV} \, , & \Omega_M  = 0.27 \, .
\end{array}
\ee

The equation of state describing the ghost's (and its partner's) dynamics depends on time as expected, and its today's value is
\be\label{eos0}
w(z=0) = w_0 \approx - 0.75 \, .
\ee

\FIGURE{\epsfig{file=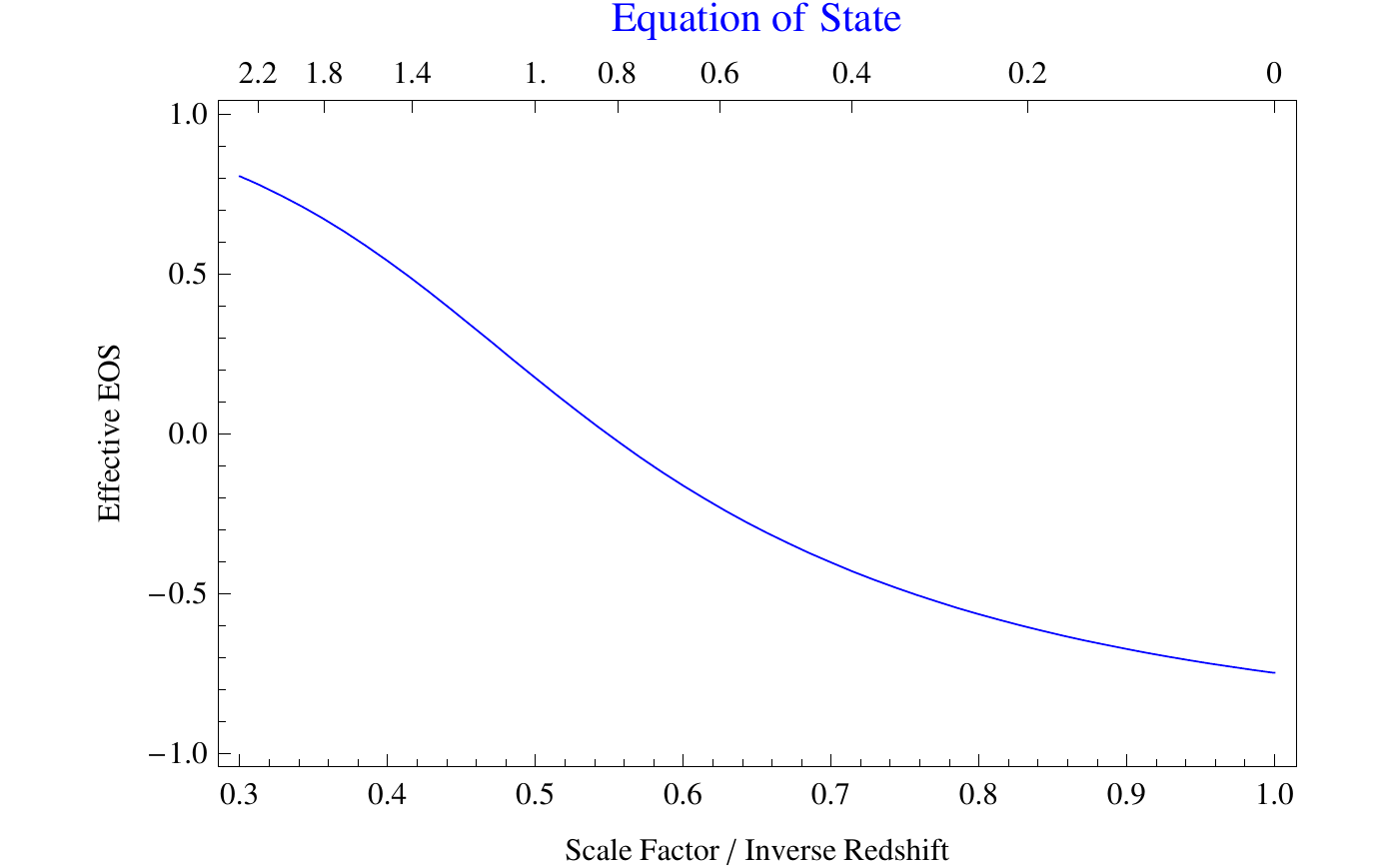}
\caption{\footnotesize{The equation of state solution of equation~(\ref{eosN}) in the text.  The upper x-axis is the actual redshift.  The equation of state is redshift (i.e., time) dependent, and today it takes the value $w_0 \approx -0.75$.}}
\label{eosPlot}}

Two facts should be mentioned concerning this result.  First of all, it is a representative result and while it is self-consistent, it should be not considered as the final word of this framework due to a number of subtleties discussed above and more deeply in appendix~\ref{Anum}.  Had we been able to compute the propagators in the compact manifold and perform the subtraction as prescribed in an entirely analytical way, the will still be freedom of variation of these two parameters.  In addition, the function $f(x)$ was introduced to supply for our inability to compute the required Green's function on a non-static compact manifold.  While its asymptotical behaviour is predictable, a precise expression for $f(x)$ is not known.  Our phenomenological parametrisation is not unique, and different $f$s conduce to similar, but still different results.  It is not our intention here to play around with different choices of $f(x)$ satisfying the same asymptotic behaviours; we instead want to present a typical, representative result with the simplest function that respects the imposed asymptotics. 

Second, the equation of state~(\ref{eos0}) is about $2\sigma$ off according to WMAP~\cite{wmap}.  However, this statement applies to an equation of state that does not evolve with time, and is significantly weakened when this possibility is taken into account.  Indeed, WMAP in this case only measures the integrated equation of state from last scattering to now, which does not provide much information on its effective time dependence~\cite{bruce}.  Moreover, the linear redshift dependence that is most widely used, that is, $w(z) = w_0 + w_a (1-a)$ does not necessarily provide a good fit for the behaviour shown in fig.~\ref{eosPlot}.

One more comment concerns the sign of $c_\phi$.  Although it has been fixed greater than zero for it is the value that best meets the criterion enunciated before, we can also find solutions to eq.~(\ref{eosN}) in the opposite case.  For instance, a solution with $w(z=0) \approx -1$ approached from below (phantom) is available for $|c_\phi| \lesssim 10^{-3}$ in the negative $c_\phi$ branch.  This, as mentioned previously, could have interesting application in models of bouncing cosmologies~\cite{bounce}.

Direct measurements of the Hubble parameter however can constrain the equation of state of dark energy directly, although such measurements are difficult to perform and inherently bear large error bars.  Using the most recent data available~\cite{hubz} we can graphically compare the standard cosmological model ($\Lambda$CDM, for Cold Dark Matter and cosmological constant) with the result of the equation of state we have just obtained.  This is shown in fig.~\ref{hubblePlot}, where it is seen how the Hubble parameter does not fit the data points.

\FIGURE{\epsfig{file=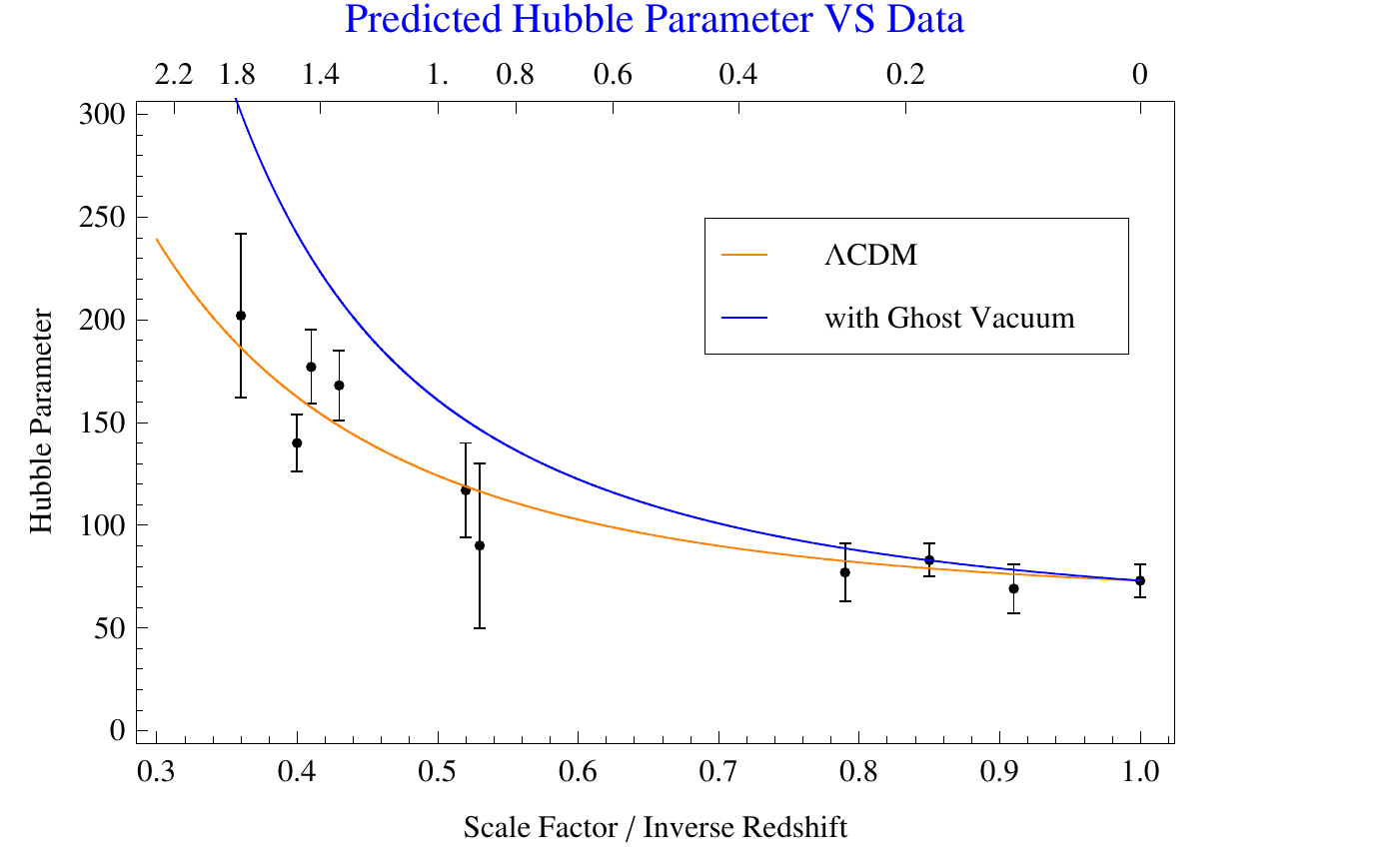}
\caption{\footnotesize{Hubble parameter as a function of the scale factor of the redshift.  The upper x-axis is the actual redshift.  As explained in the legenda, the two lines correspond to the $\Lambda$CDM model (orange, or light grey in black/white) and the model proposed in this work (blue, or dark grey in white/black).  The data are given with the corresponding error bars.}}
\label{hubblePlot}}

The last graphical result we want to include is the behaviour of the different components of energy density of the universe for the same redshift range for which direct measurements of the Hubble parameter are available.  In figure~\ref{rhoPlot} we show the evolution of the vacuum energy component described in this work, together with the flat cosmological constant and the total matter energy density.

\FIGURE{\epsfig{file=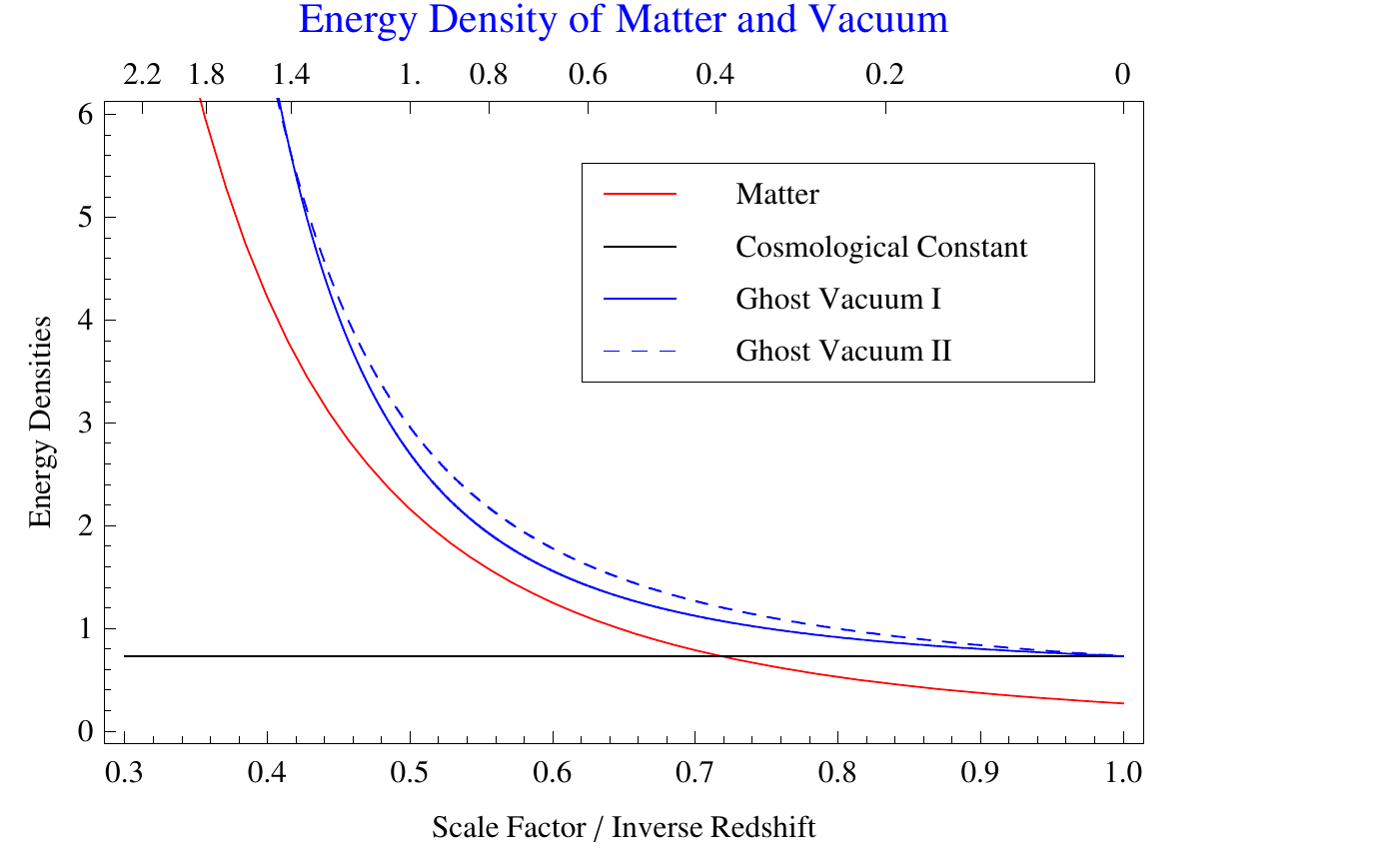}
\caption{\footnotesize{Energy densities against scale factor.  The upper x-axis is the actual redshift.  As shown in the legenda, the solid blue (dark grey in black/white) line tracks the vacuum energy density using expression~(\ref{rhoI}), while the dashed line follows~(\ref{rhoII}). The flat black line is a constant cosmological constant, and the red (light grey in black/white) one is the energy density of total matter.}}
\label{rhoPlot}}

Obviously, the vacuum energy density in our framework is no longer constant, and in principle it will meet that of ordinary and not-so-ordinary matter at earlier times than usually assumed based on $\Lambda$CDM model. However, the classical solutions we have found seem to point to an energy density which at early times grows faster than matter or radiation, thereby overwhelming them.  With the choice endorsed in~(\ref{const}) we also find a model in which matter never dominates; a much smaller value for $c_\phi$ would have returned a better agreement with observations, although mining the consistency of the different expressions for the energy density (see below).  For instance, we can obtain an equation of state virtually indistinguishable from a constant $w \simeq -1$ in the desired redshift range for $c_\phi \lesssim 10^{-6}$, which would re-establish a viable cosmology.  A more serious problem appears to be the too fast growing of the kinetic energy density of the ghost in the past universe; this feature as well should not represent a too serious menace for this proposal, as it crucially depends on the assumptions we have made on the shape and time dependence of the potential.

In general however, our results suggest that there will be appreciable amount of dark energy even at very high redshifts ($z>5$) because of the specific features of the equation of state derived above. This is in contrast with the standard $\Lambda$CDM model where, in the same redshift region, the vacuum energy is completely ignorable.  The study of this redshift span ($z>5$) using Gamma Ray Bursts (GRB) as standard candles as suggested by ref.~\cite{Hooper:2005xx} (see also the analysis of~\cite{Basilakos:2008tp} and references therein) represents a unique opportunity to verify or rule out our proposal.  Indeed, as the mechanism described in this paper unambiguously predicts sizeable amounts of dark energy at very high redshifts, one can discriminate between models by studying $z>5$ (see also the recent work~\cite{Xia:2009ys} on constraints on dark energy contribution for $z \simeq (2-20)$.  Finally, let us notice that a joint likelihood analysis of three different types of observations (supernovae type Ia data, CMB, and the Baryonic Acoustic Oscillations), similar to what has been lately performed in~\cite{Basilakos:2009wi}, is also sensitive to the peculiar features of our equation of state~(\ref{eos}).

Coming back to fig.~\ref{rhoPlot}, notice that there are two lines corresponding to the vacuum energy because there are two possible ways of expressing it, either based on the parametrisation through the effective equation of state
\be\label{rhoI}
\rho_\Lambda(z) = \rho_\Lambda(0) e^{-3 \int \!\dd a (w+1)/a} \, ,
\ee
with $w(a)$ as obtained graphically from the solution of equation~(\ref{eosN}) (see appendix~\ref{Anum}), or directly from the solution of the equations of motion for the ghost
\be\label{rhoII}
\rho_\Lambda(z) &=& \alpha_1 H(z) (1+z)^3 + \alpha_2 (1+z) f(x) \, ,
\ee
where $f(x)$ is defined by eq.~(\ref{fx}) and exhibits a very non-trivial dependence on $z$.  As inferred from the plot, with the choice of $c_\phi$ and $c_g$ in eq.~(\ref{const}) these two expressions are consistent within 15\%.  Let us stress that some mismatch between them (which are supposed to be exactly identical) is not a result of some fundamental flaw, but rather springs from our inability to compute $f(x)$ analytically, combined with the fact that we are forcefully imposing a potential which has been derived in a purely quantum mechanical way (and that is a result of non-trivial boundary conditions) into the classical system.  Therefore, these numerical results should not be treated as the final result: the full complete quantum calculation would return expressions that are automatically consistent one another, although their graphical representation by numerical means (recall that $H$ and therefore $w(a)$ itself appear in the expression for $w = w(a)$ eq.~(\ref{omega})) will introduce once more some degree of numerical mismatch.

\subsection{The fate of the universe}

The evolution of the universe can be followed beyond today thanks to the fact that we have a precise equation telling about the behaviour of the vacuum energy component at any time in the future.  Indeed, one can look at the expressions for the equation of state, which, once a specific form for $f(x)$ is selected, leads immediately to the evolution of the vacuum energy density.  Looking deep into the future, the equation of state is going to be
\be\label{eosDoom}
w(a) = \frac{H(z) + \alpha a^2 f(x)}{H(z) - \alpha a^2 f(x)} \rar \frac{1 + \alpha a^3}{1 - \alpha a^3} \rar -1\, .
\ee
Thus, the equation of motion (for this $f(x)$) will slowly approach -1 from above, and the universe is dragged into a de Sitter state, just as in the usual $\Lambda$CDM model.

Indeed, the energy density of vacuum will more and more approach a constant value, result that can be seen either in the parametrisation~(\ref{rhoI}) when $w(a) \rar -1$ or in~(\ref{rhoII}) since in this case the dominant piece will be the potential energy as given in~(\ref{potentialf}) which when $f(x) \rar x$ becomes proportional to $H$, which itself is approaching a constant.  This shows that asymptotically the form chosen for $f(x)$ is self-consistent.

\section{Fine tuning without fine tuning}\label{tuning}

This section aims at explaining in a non-technical, intuitive way how a number of fine tuning issues such as ``coincidence problems'', ``drastic separation of scales'', ``unnatural weakness of interactions'', etc., which always plague dark energy models~\cite{ed,turner}, possess simple explanations within the framework endorsed in this paper, even without the need to introduce new fields, which come with new interactions, new coupling constants, and new symmetries.

$\bullet$ We start with a close examination of the energy density scale $\rho_{\Lambda} \approx (2.3\cdot 10^{-3} \text{eV})^4$: where does it come from in our set up?  Basically, this scale emerges as a result of our operational definition for the physical dark energy.  Indeed, according to the scheme proposed in the introduction, the ``renormalised cosmological constant'' is set to be zero in Minkowski space wherein the Einstein equations are automatically satisfied.  Thus, the energy-momentum tensor in combination with this ``bare cosmological constant'' must also vanish at this specific ``point of normalisation'' to satisfy the Einstein equations.  From this definition it is quite obvious that the ``renormalised energy density'' must be proportional to the departure from Minkowski spacetime geometry.

In the context of QCD on a compact manifold with characteristic size $L$, this extra piece gives precisely the appropriate scale for $\rho_{\Lambda}$ as an offspring of known QCD dynamics at the QCD scale~\cite{our4d}. One should emphasise that this contribution is unique as all other vacuum contributions (such as gluon or Higgs condensates) vanish due to their cancellation with the vacuum energy defined in Minkowski spacetime, or are exponentially suppressed.

$\bullet$ The second fine tuning problem goes under the name of ``cosmic coincidence'', or: why does it happen now?  The same type of arguments which have been presented above suggests that the ghost vacuum energy should become relevant when its potential energy is of the same order of magnitude as $\rho_c$.  Equating these two quantities returns
\be
\label{V}
V \simeq \frac{2 N_f  |m_q\la\bar{q}q\ra  |}{m_{\eta'}L_0} \simeq 3H_0^2 M_P^2 \, ,
\ee
which can be read as an estimate for the age of the universe as $t_0 \simeq H_0^{-1} \simeq 10h^{-1}$Gyr, which is indeed the correct lifetime for the present Hubble size.

$\bullet$ Most of the proposals on how to attack the vacuum energy problem suggest to treat dark energy using a scalar field $\phi$ (``quintessence'', ``K-essence'', etc., see the reviews~\cite{ed,turner}).  The potential describing the dynamics of this new field $\phi$ must be extremely flat, with a typical mass $m_{\phi}\simeq 10^{-33}$eV~\cite{ed,turner}.  Such a tiny number is unheard of in particle physics, and certainly requires a great degree of fine tuning.

In our model it so happened that the dynamics is also determined by interacting scalar fields, see section~\ref{ghostlagrangian}.  However, while the interactions among them is strong (it comes from strongly interacting QCD), the potential energy which remains after the subtraction of the large Minkowski piece is indeed very shallow~(\ref{V}), and becomes even more shallow with time since the potential energy scales as $1/L \simeq 1/a$.  The moral is that the effective potential is indeed flat, but this flatness does not stem from some approximate original symmetry (which is typically exploited in dark energy proposals~\cite{ed,turner}), but rather is a consequence of the same governing principle stating that the physical energy density must be proportional to the \emph{deviation} from Minkowski spacetime geometry.  This difference in our case is ${\cal O}(1/L)$.

There are no new particles in our framework.  Instead, there are the Veneziano ghost and its partner which are known low-energy fields of QCD and are responsible for the solution of the $U(1)_A$ problem.  These fields are always present in the conventional QCD in Minkowski space.  Their dynamics is trivial in Minkowski spacetime, and was therefore ignored in all previous QCD applications.  The scenario is analogous to that of the Gupta-Bleuler quantisation of QED~\cite{G,B}, where photons with  time and longitudinal polarisations can be ignored in all S-matrix elements, because in the gauge invariant sector they always cancel each other. The ghost and its partner experience a new life in a cosmological background, and lead to time-dependent vacuum energy in curved space QCD.  They are still not physical (i.e., asymptotic) states.  Nevertheless, their contribution to the vacuum energy is different from its Minkowski couterpart, and can be thought of as a time-dependent ``ghost condensation''.

Finally note that QED photons, including unphysical polarisations, may also in principle contribute to dark energy.  This contribution however is very small, as it is of order of $L^{-4}$ or $H^4$ by dimensional reasons (see however the mechanism proposed in~\cite{qed1,qed2}).

$\bullet$ Not only well behaving scalar fields, but also ghost fields (similar to the Veneziano ghost with a negative sign for the kinetic energy term) have been extensively discussed in order to describe dark energy~\cite{ed,turner}.  Unfortunately, they normally lead to more questions than answers.  As is known, ghost fields (``phantom fields'' in the context of cosmology) are generally pestered by severe quantum instabilities, violating unitarity and other crucial fundamental principles of QFT.  Why does our system not suffer from all these severe problems?  The answer lies in the existence of the $\phi_2$ field, which always pairs up with the Veneziano ghost $\phi_1$ in QCD, see eq.~(\ref{lagrangian}).  These two unphysical degrees of freedom $\phi_1$ and $\phi_2$ drop out of every gauge invariant matrix element in Minkowski space, leaving the theory well defined, i.e., unitary and without negative normed physical states after the subsidiary condition~(\ref{gb}) or~(\ref{gb1}) is imposed.

As we argued in the previous section however, when QCD is defined on a curved background, the exact cancellation between $\phi_1$ and $\phi_2$ does not hold anymore, and these fields do contribute to the energy~(\ref{H2}).  However, as we claimed, the departure from Minkowski space is controlled by $1/L\sim H$ rather than by the expected QCD scale $\Lqcd$.  This stems from the exact cancellation between $\phi_1$ and $\phi_2$ that takes place in Minkowski spacetime.

In particular, the constraint on the ghost maximal energy ($k_{max}\lesssim 20$ MeV) derived in~\cite{Cline:2003gs} is obviously satisfied.  Indeed, as we have already elaborated on previously, the energy which can be associated with the ghost in the expanding universe is only $\omega_k \simeq k \simeq H$.  Therefore, since $H \sim 10^{-33} \mathrm{eV} \ll k_{max}$, we clearly meet the limit of~\cite{Cline:2003gs} on possible dynamical violations of unitarity.  Let us mention once more that this very low cutoff comes about because of the subsidiary condition~(\ref{gb}) (or, equivalently,~(\ref{gb1})) on the physical Hilbert space, and hence, it is a \emph{result} of the dynamics of the ghost in curved space, rather than an \emph{ad hoc} requirement on the form of the ghost's interactions.

$\bullet$ Last, but not least, the same feature of the spectrum discussed above allows us to name this extra ghost dipole contribution~(\ref{H2}) dark energy; indeed, since the typical wavelengths $\lambda_k$ associated with this energy density are of the order of the Hubble parameter, $\lambda_k\sim 1/k\sim 1/H\sim 10\textrm{~Gyr}$, the corresponding excitations are not going to produce any clustering on scales smaller than this, as opposed to visible or dark matter.

\section{Conclusion}\label{conclude}

Dark energy has provided formidable headaches to physicists for over a decade. In trying to understand the physical principles and mechanisms, sapiently hidden by Nature, that are responsible for it, the physics community has produced a mole of models that is quickly climbing up to 4000 arXiv papers.  On the experimental side, a great deal of efforts is being devoted in planning, projecting and performing tests which will help theorists pierce the veil of Maya of dark energy.

In our previous paper~\cite{our4d} we have claimed that 4d QCD, once formulated on a compact manifold, will exhibit a very peculiar feature, that is, the vacuum state will depend \emph{linearly} on the inverse characteristic size $L$ of the manifold.  This finding was supported by an explicit calculation in the context of the 2d Schwinger model~\cite{toy}, where the linear dependence can be tracked analytically all the way to the final result.  The non-vanishing result~(\ref{rhov}) can be understood as a Casimir-type of vacuum energy when the boundary conditions and topology play a key r\^ole.

The essential reason for this unexpected behaviour is to be traced back to the existence of a very special degree of freedom, the Kogut-Susskind  ghost in 2d and the Veneziano ghost in 4d, which, despite being unphysical, are able to carry the information about the boundaries of spacetime due to its protected masslessness.  Once the existence and the form of the QCD ghost's potential vacuum energy is established to be~(\ref{rhov}), the next logical step is to promote the background to a general curved background, in our case a FLRW spacetime, and study the dynamics of the system which is  precisely  the  goal of the present work.  The main achievements of these studies are listed below.\\

{\bf 1.}  The Veneziano ghost $\phi_1$ is always accompanied by its partner $\phi_2$.  These two unphysical degrees of freedom $\phi_1$ and $\phi_2$ drop out of every gauge invariant matrix element in Minkowski space, leaving the theory well defined, i.e., unitary and without negative normed physical states after the subsidiary condition~(\ref{gb}) or~(\ref{gb1}) is imposed.  The only place where the Veneziano ghost manifests itself is the topological susceptibility, because the ghost's partner $\phi_2$ does not participate in its expression.  However, it is through the topological susceptibility that we know about the existence of the ghost and its properties.\\

{\bf 2.}  This simple picture drastically changes when QCD is coupled to gravity, in which case another physically observable manifestation of the ghost emerges.  In this case the ``would be'' unphysical ghost contributes a physical portion of the energy density of an expanding universe.  All effects related to this contribution are proportional to the rate of expansion $H$ such that numerically they are naturally very small, $H/ \Lqcd \sim 10^{-41}$.  However, this ``tiny'' effect is precisely what accelerates our universe today.  In the limit for which $H \rar 0$, which corresponds to the usual Minkowski QFT formulation, every additional contribution vanishes.  We interpret this extra contribution as a time-dependent ``ghost condensation'' as explained at the end of section~\ref{curved}.  We are not claiming that the ghost field becomes a propagating degree of freedom, or becomes an asymptotic state.  Rather, we claim that our description in terms of the ghost is a convenient way to account for the physics hidden in the non-trivial boundary conditions as discussed in section~\ref{wits}.  The physical phenomena (described by the Veneziano ghost in our framework) do not disappear when we use a different approach, for instance when the ghost is not even a part of the system, see section~\ref{wits}.  However, we do not know presently how to track this physics in a ghost-free formulation.  We strongly suspect that the corresponding description (if found) would be much more technically complicated in comparison with the approach presented in this work.\\

{\bf 3.}  The usual catalogue of fine tuning issues which typically pester us while tampering with dark energy does not appear here, as these questions possess simple explanations in our scenario, without the need to introduce new fields, which come with new interactions, new coupling constants, and new symmetries.  The nature of this ``fine tuning without fine tuning'', see section~\ref{tuning}, is not to be sought for in supersymmetry or any other extra symmetries imposed on the system (there are in fact, none), but it comes about from the auxiliary conditions on the physical Hilbert space~(\ref{gb}) which accommodate the gigantic span of scales ($\Lqcd$ versus $H_0$).\\

{\bf 4.} A word on the testability of our framework.  There are in fact several places where one can look for supporting or ruling out evidence for our proposal.  First of all, as we have claimed in~\cite{our4d}, the linear dependence of the vacuum energy on the inverse size of the manifold can be tested by studying the topological susceptibility of QCD on the lattice by looking for its dependence on $L$.  Second, we have argued in~\cite{cmbt} that the size of the manifold is not overly large, and could in principle be tested in upcoming CMB maps (PLANK). Third, since in this paper we have provided the dynamical equations that govern the evolution of the ghost, its contribution to the Hubble parameter and to the expansion of the universe in general, its equation of state $w(t)$ and its effects on structure formation, can all be confronted with cosmological observations (such as supernovae type Ia data, CMB, the Baryonic Acoustic Oscillations, Gamma Ray Bursts, see section~\ref{results} for details), a feature not often found in dark energy models.

Finally, we would like to draw attention to the fact that it is quite amazing that QCD, which na\"ively has no relation what so ever to the dark energy problem, may in fact be indeed the locus one should be scrutinising to uncover the dark energy secret, as we argued in this work.  The emergence of the QCD scale in the vacuum energy may eventually unlock the mystery of the ``cosmic coincidence'' problem, that is, the observational similarity among visible and invisible scales: $\Omega_{\Lambda} \simeq 4 \Omega_{M}$ and $\Omega_{M } \simeq 5 \Omega_{B}$ where only $\Omega_{B}$ (which represents the baryon contribution to $\Omega_{tot}=1$) has an obvious relation to QCD, as the nucleon mass $m_N$ is proportional to $\Lqcd$, while ``God's particle'' (the Higgs boson), contributes only few percents to $m_N$ (and correspondingly to $\Omega_{B}$) through the quark's mass dependence on the Higgs vev.  It is also fairly interesting that a number of dark matter puzzles (along with the dark energy mystery which is the subject of the present study), may also be related to QCD~\footnote{We refer to few recent papers on the subject~\cite{Zhitnitsky:2006vt, Forbes:2008uf} and references therein, where it has been argued that dark matter, in fact, may also have QCD origin.  The basic idea is that nuggets of very dense matter similar to Witten's strangelets~\cite{Witten:1984rs} form at the same QCD phase transition as conventional baryons (neutrons and protons), providing a natural explanation for the similarity between $\Omega_{M}$ and $\Omega_{B}$~\cite{Oaknin:2003uv}.}.  The different sides of our construction allow for several types of tests to be performed, and some of them are radically new (and are entirely intertwined with the QCD nature of dark energy).  Ultimately, observations will be able to pin down the winner, the model which most adherently follow the forms of nature.

\section*{Acknowledgements}

The authors would like to acknowledge Doug Scott, Robert Brandenberger and Federico Piazza and the participants of the Emergent Gravity IV workshop held at UBC for stimulating discussions.  We would like to thank Cliff Burgess who suggested that we explicitly mentioned QED in this work.  Moreover, FU thanks Paul Laurain for spending an entire day explaining mathematics to a physicist.  This research was supported in part by the Natural Sciences and Engineering Research Council of Canada.

\appendix

\section{Numerics}\label{Anum}

This first appendix outlines some details of the numerical calculations which produced our figures.  The first quantity we want to compute is the equation of state of the vacuum energy fluid.  This is given by equation~(\ref{eosN}) in the text which, once we incorporate the effects of the unknown function $f(x)$ becomes
\be\label{eosA}
w(a) = \frac{\left[ \Omega_M^0 a^{-3} + \Omega_\Lambda^0 e^I \right]^{1/2} - \alpha a^2 f(x)}{\left[ \Omega_M^0 a^{-3} + \Omega_\Lambda^0 e^I \right]^{1/2} + \alpha a^2 f(x)} \, ,
\ee
where, following the main text, we choose $f(x) = 1 + x$ and where $x = HL$.  It is not possible to solve for $w(a)$ analytically, as it appears both on the l.h.s.\ and on the r.h.r.\ under the integral
\be\label{integr}
I \equiv -3 \int \!\dd a \frac{1+w(a)}{a} \, .
\ee

The easiest way to numerically overcome this problem is to assume that in the integral the equation of state is a constant, and then solve graphically for $w(a)$.  Of course the resulting equation of state will not be a constant, but it can be used back in the integral~(\ref{integr}) which can be solved numerically and the resulting expression plugged back into~(\ref{eosA}) to obtain (once more, numerically) a better approximation for $w(a)$.  This is the procedure followed in obtaining the results shown in the figures in the text, figs.~\ref{eosPlot},~\ref{hubblePlot}, and~\ref{rhoPlot}.

Practically, we have found that within two recursions we already have a reliable approximation which does not change much employing further steps.  The reason for this can be traced back to the fact that among the two components of energy and pressure density, that is, kinetic and potential term, at low redshift the potential term dominates (for our choice of the parameters) and the errors induced by mistaking $w$ for constant in $H$ mainly appear in the kinetic term.  In the high $z$ region, when the kinetic term dominates, there will be some more uncertainty, but this stabilises within a couple of steps.

Once a plot for the equation of state has been obtained (which comes from feedbacking $w$ into~(\ref{eosA}) once) we can use an interpolation of this to find the other quantities of interest, in particular the Hubble parameter and the energy density of the vacuum.

As mentioned in the main text, since we are forced to introduce the function $f(x)$ together with the fact that the information about the compact manifold is implicitly hidden in some of our expressions while it has been entirely discarded in others, will introduce some degree of inconsistency among different expressions for the same quantity.  The clearest example of this can be immediately noticed by looking at the expression for the vacuum energy as given in equation~(\ref{rhoII}): the potential term comes from a very specific subtraction procedure as explained in~\cite{our4d} and exhibits the $1/L$ behaviour which is one of the principal ingredients of our work.  This subtraction has been defined and performed in a compact Minkowski space, and if promoted acritically to an expanding universe it would lead, in the case of a simple 3-torus with the same expansion rate in each direction, to the behaviour $V_{\mathrm{ghost}} \propto 1/a$, which implies an equation of state $w = -2/3$, at least in the redshift range for which the potential energy dominates over the kinetic energy.  At the same time the equation of state clearly shows that if this is the case then the equation of state approaches -1, in contradiction to what has been found just a few lines above.

The resolution to this apparent paradox lies in the subtleties mentioned in the text and analysed here, that is, the expression~(\ref{potential}) is based on our understanding of the ghost dynamics without gravity: it can be analytically computed in the 2d Schwinger model and hopefully it can be tested in 4d QCD using the lattice QCD computations as explained in~\cite{our4d,toy}.  This contribution to the vacuum energy is computed using QFT techniques in a non-expanding universe.  As it stands, it can not be used for studying its evolution with the expansion of the universe.  In order to do so one needs to know the dynamics of the ghost field coupled to gravity on a finite manifold, which we are not able to solve for analytically, and we therefore must resort to modelling it by means of the function $f(x)$.

Even doing so there will still be some inconsistency due to the fact that the information about the compact manifold that shows up in the potential energy indirectly influences also the resulting kinetic energy.  However in this case the effects are subtler and it is not possible to easily model them as for the potential term.  Our approach in this case is to use the freedom of choosing the two parameters $c_\phi$ and $c_g$ to minimise the impact of these uncertainties on the quantities of interest, which will leave us with a consistent \emph{representative} theory.  This theory is not unique and may well not be able to fully reproduce the observational data, see the discussion in the main body of the paper.

Thus, the operational criterion for the choice of the constants~(\ref{const}) is that the energy density of the vacuum computed using the effective equation of state or using directly the solutions for the equations of motion (formulas~(\ref{rhoI}) and~(\ref{rhoII}), respectively) must be the same within 15\%. If one fixes the energy density of the vacuum today to $\Omega_\Lambda^0 = 0.73$ without errors, this automatically fixes one parameter in terms of the other as
\be\label{cs}
\alpha_2 = \frac{\rho_\Lambda^0 - \alpha_1}{f(1)} \, ,
\ee
which in turn gives $c_g$ in terms of $c_\phi$.

\FIGURE{\epsfig{file=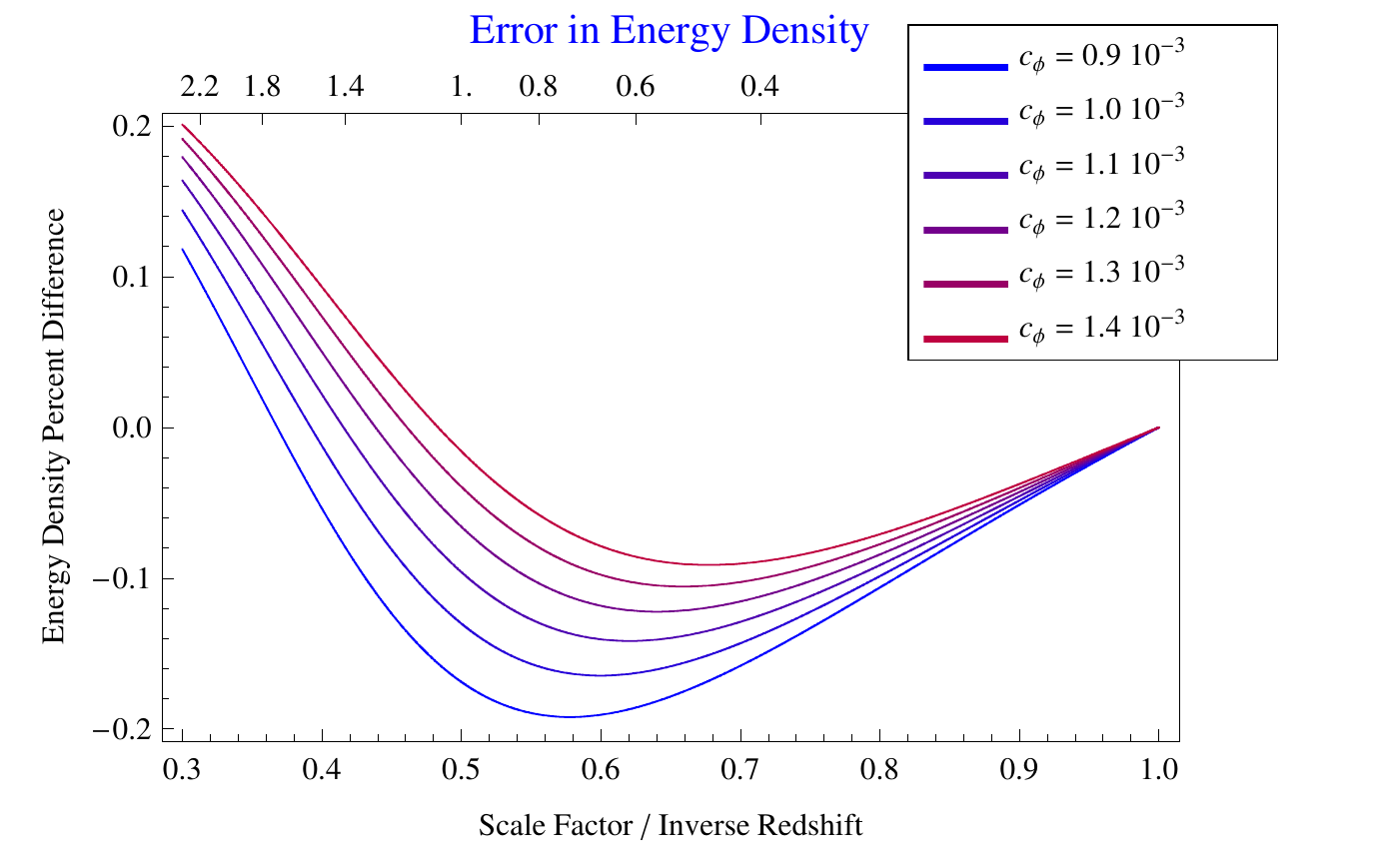}
\caption{\footnotesize{Difference in percent between the two parametrisations of the vacuum energy density for different values of $c_\phi$ as a function of the scale factor.  As usual, the upper x-axis is the actual redshift.}}
\label{errPlot}}

This criterion led to the choice~(\ref{const}), and will be different for a different choice of the unknown $f(x)$.  In figure~\ref{errPlot} one can see how by moving away from the choice values the two expressions~(\ref{rhoI}) and~(\ref{rhoII}) progressively drift farer apart.  Allowing for an error in the determination of the vacuum energy density today would open up some more parameter space where the criterion formulated above would be satisfied.  Finally, notice that the resulting $c_g$ is slightly different than the one inferred in our paper~\cite{cmbt} but only marginally ($c_g \approx 0.059$ in that paper, whereas $c_g \approx 0.051$ here), since at zero redshift most of the vacuum energy density is in fact stored in the potential energy, fact that confirms the validity of the hypothesis made (and the results thereby obtained) in that work.

\section{Compact universes}\label{Atorus}

The idea of a compact universe has surfaced in several places throughout the paper.  It is only thanks to the embedding of the spatial hypersurfaces in a compact 3d manifold that we expect to observe a linear departure from the usual vacuum contributions in the form of $H^2$ or $H^4$.  Moreover, although in a much more subtle way, the linear dependence of the potential energy as in~(\ref{potential}) or~(\ref{potentialf}) on the characteristic size of the compactified manifold also affects the solutions for the differential equations and therefore the kinetic piece~(\ref{kinetic}) of the total energy density of the Veneziano ghost.

In the main body of the paper we have been rather generic in discussing this embedding in a compact manifold, the reason for this being the fact that in this way we could obtain interesting, self-consistent results with the minimal amount of free parameters (in fact, only one, $c_\phi$).  Consequently, taking into account the more general case of an anisotropic homogeneous expanding compact manifold would only enlarge the space of possibilities, however at the expense of clarity.  Yet one more motivation for not carrying out the calculations in the most general set up is given by the acknowledgement that the most paramount quantity in our work, the value of the chiral condensate in a given background, is not known (and discouragingly hard to calculate), thereby eluding a completely closed analysis.

For the sake of completeness, we will nevertheless review some of the key concepts and results in compact 3d geometries in this appendix.  A clear and thorough treatment of the topic can be found in~\cite{jl,maths1,maths2,maths3}.  Let us begin with a few definitions (in this appendix we employ the abstract index notation as in~\cite{wald}).  We are interested in ``geometries'' in the sense of a pair, $(^{(4)}\tilde M , \tilde g_{ab})$, where $^{(4)}\tilde M$ is a non-compact simply connected Lorentzian manifold, and $\tilde g_{ab}$ its metric, or universal cover, a solution of the Einstein equations.  This geometry needs to be reduced to a 3d Riemannian one that we call $(\tilde M , \tilde h(t)_{ab})$, at each given time (for the purpose of engineering a compact 3d slice of the 4d manifold the time lapse and shift functions can not depend on the spatial coordinates, but only on time itself), whose compact subgeometries can be studied.  Here the term ``subgeometry'' refers to the fact that a geometry $(X , h_{ab})$ can always be thought of as a pair $(X , H)$ where $H$ is some symmetry group, and the corresponding $H$-invariant metric can be always constructed: therefore a pair $(X , Q)$ with $Q\subset H$ is a subgeometry of $(X , H)$, and the same applies to the corresponding manifold-metric pairs.  In this example $H$ would be the group of the isometries of the metric $h_{ab}$.

In order to make the 3d manifold compact, and find the corresponding metric, we need to know the group of isometries of $(\tilde M , \tilde h_{ab})$, which we denote $\mathrm{Isom} \tilde M$ (in fact, we need the subgroup of $\mathrm{Isom} \tilde M$ which is orientation-preserving and extendible).  Once this group is identified we make $(\tilde M , \tilde h_{ab})$ compact by taking its quotient with a discrete subgroup of $\mathrm{Isom} \tilde M$, which we name $\Gamma$.  This group is basically the group of the identifications that wrap and glue the manifold to make it compact, and it is referred to as the covering transformations group.  Therefore, the resulting compact 3d geometry $(M , h_{ab})$ is metrically diffeomorphic to $(\tilde M , \tilde h_{ab}) / \Gamma$.  At this point one can jump back to 4d and naturally extend the covering group $\Gamma$ to the Lorentzian geometry $(^{(4)}\tilde M , \tilde g_{ab})$: the resulting compact 4d spacetime $(^{(4)}M , g_{ab})$ will inherit the metric from the universal cover $\tilde g_{ab}$, and if that is homogeneous, so will the compact space metric $g_{ab}$ be.

Notice that thinking of a compact space as of $\tilde M / \Gamma$ automatically separates the degrees of freedom of $M$ to those of $\Gamma$.  The latter admits a group of smooth deformations (for instance, stretching or skewing a parallelepiped in 2d) which are known as the Teichm\"uller deformations, which form a group and are parametrised by the Teichm\"uller parameters.  Therefore, the Teichm\"uller parameters know about the global structure of the manifold while preserving all local quantities, whereas the local coordinate parameters are insensitive to the global structure.  In general there will be degrees of freedom associated with the Teichm\"uller parameters, and they may have some dynamical r\^ole, which for simplicity we neglect in our report.  The complete classification of the possible compact topologies and their Teichm\"uller parameters is given in~\cite{maths1,maths2,maths3}.

With all the definitions at hand, we can present a specific example for the simplest compact manifold in a flat (i.e., Euclidean) 3d space, that is, the three-torus $T^3$.  In general this type of compactification is admitted in the context of the Bianchi I or Bianchi VII(0) geometries only.  These geometries of course admit also other more complicated orientable compactifications, e.g., $T^3 / \mathbb{Z}_2$ or $T^3 / \mathbb{Z}_3$ etc., each of which will come with its own Teichm\"uller space.  For concreteness, we stick to a non-twisted $T^3$ on Bianchi I geometry.

The aim is to find the most general metric which contains all the information about the dynamical local variables and the global deformations of the manifold.  Schematically, the process goes as follows.  First of all, we identify the 4d universal cover, that is, we specify $\tilde g_{ab}$ from which we immediately derive the 3d slice $(E^3 , \tilde h_{ab})$.  This is going to be a particular solution of the Bianchi I type, which for instance, once a coordinate basis has been chosen, is given by
\be\label{bianchiI}
\dd s^2 = \dd t^2 - a(t)^2 \dd x^2 - b(t)^2 \dd y^2 - c(t)^2 \dd z^2 \, ,
\ee
with three, in general unequal, scale factors.

At this point one can choose a specific 3-torus $T^3_g$ by giving its fundamental domain.  The resulting geometry will therefore be $(T_g^3 , H_{ab})$ where $H_{ab}$ is the metric inherited from the universal cover $\tilde h_{ab}$ (in fact, it is the same).  However this is not yet what we want, as the degrees of freedom corresponding to the Teichm\"uller deformations are hidden in the definition of the compact manifold, that is, in its covering group.  It is however always possible to transfer the information about the global structure of the manifold to the metric.  This can be achieved by pulling back the pair $(T_g^3 , H_{ab})$ to $(T_0^3 , h_{ab})$, where now $T_0^3$ is the reference 3-torus (the unit cube), and $h_{ab}$ contains all the global and local geometrical information.

In order to calculate the metric $h_{ab}$ we need the pullback from the torus $T^3_g$ to the reference $T^3_0$ one, which can be written explicitly in terms of the generators of the group of the isometries defining the covering group, which in this case are three dimensional rotations and translations, parametrised by the three three-vectors $g_1$, $g_2$, and $g_3$.  Precisely how this is done is detailed in~\cite{maths1,maths2,maths3}, and we only repeat their results here.  In the Bianchi I case given in~(\ref{bianchiI}) the result would be
\be\label{metrich1}
h_{ij} = a(t)^2 g_i^1 g_j^1 + b^2(t) g_i^2 g_j^2 + c(t)^2 g_i^3 g_j^3 \, .
\ee

It is easy at this point to make the connection with the Teichm\"uller parameters.  Indeed, every metric on the reference torus $T_0^3$ can always be parametrised in terms of the Euclidean metric suitably deformed by the Teichm\"uller group.  This is to say, we can always describe $(T_0^3 , h_{ab})$ in terms of $(T_{\mathrm{Teich}}^3 , \eta_{ab})$, where $\eta_{ab}$ is the Euclidean 3d metric and $T^3_{\mathrm{Teich}}$ is the deformed torus.  The advantage of using the Teichm\"uller parameters comes from their easy geometrical interpretation in terms of global distortions of a compact manifold, and that this scheme can be implemented for each and every possible compact space of a general homogeneous (although not necessarily isotropic) metric.

The specific shape of these distortions is thus going to be parametrised by the Teichm\"uller parameters, that for the 3-torus can be thought of as three-vectors identifying the axes of the manifold.  Of course we do not need nine parameters to describe the manifold, as we can always choose the first vector to lie on the, say, $x$-axis, and the second vector to span the $xy$-plane.  In this case the most general deformation can be obtained by the following generators
\be\label{teich}
a_1 = \left( \begin{array}{c} a_1^1 \\ 0 \\ 0 \end{array} \right) \; , \;
a_2 = \left( \begin{array}{c} a_2^1 \\ a_2^2 \\ 0 \end{array} \right) \; , \;
a_1 = \left( \begin{array}{c} a_3^1 \\ a_3^2 \\ a_3^3 \end{array} \right) \, ,
\ee
acting on the standard Euclidean metric $\dd l^2 = \dd \bar x^2$, where $\dd s^2  = \dd t^2 - \dd l^2$.  The overall volume of the manifold is not fixed, therefore there are six degrees of freedom in the Teichm\"uller space.  The resulting metric is of course
\be\label{metrich2}
h_{ij} = \left( \begin{array}{ccc}
(a_1)^2		& a_1^1 a_2^1					& a_1^1 a_3^1 \\
a_1^1 a_2^1	& (a_2)^2						& a_2^1 a_3^1 + a_2^2 a_3^2 \\
a_1^1 a_3^1	& a_2^1 a_3^1 + a_2^2 a_3^2		& (a_3)^2
\end{array} \right) \, .
\ee

The time-dependence of the Teichm\"uller parameters can be read off by comparing the two metrics~(\ref{metrich1}) and~(\ref{metrich2}).  Of course in simple cases such as the one just worked out, it is straightforward (and intuitive) to separate the time dependence from what one calls Teichm\"uller parameters, in such a way that the metric $h_{ij}$ shows explicitly the separation in local (time-dependent) and global quantities.

As a final example, and because it is relevant for our discussion in the main text, we shall specialise further to the usual isotropic FLRW universe, setting equal the three scale factors $a(t) = b(t) = c(t)$.  If we moreover demand that the angles of the fundamental domain are all right, this leaves only diagonal Teichm\"uller parameters.  The most general 3d metric for this universe is given by
\be\label{metrich3}
\dd l^2 = a(t)^2 \left( (a_1^1)^2 \dd x^2 + (a_2^2)^2 \dd y^2 + (a_3^3)^2 \dd z^2 \right) \, ,
\ee
where we have explicitly factored out the scale factor.  This compact Bianchi I universe has a common expansion rate, but different physical sizes in the three directions of space, and therefore will not support an isotropic perfect fluid but rather will have a direction dependent equation of state~\cite{anis1,anis2}.  This means that our parametrisation of the equation of state~(\ref{eosN}) is incomplete, and that in general more freedom is allowed when the universe is compact.  In the most general case there will also be off-diagonal terms describing skew toruses, which will carry even more parameters in the attempt to describe the system.

A final comment on the relevance of the global parameters in the real world.  First, as it has just been explained, these parameters may impose some restrictions on the specific form of the energy-momentum tensor that one can employ; second, as it is well known a QFT on a compact manifold is not the same as on a non-compact one, and it is in general sensitive to the specific structure of the compact manifold through the boundary conditions imposed on the quantum fields.  We have given in~\cite{toy} a concrete example on how the \emph{physics} can be radically different when one quantises on a finite manifold rather than on infinite spacetime, where corrections even linear in the size of the fundamental compact domain may arise in physical quantities such as the topological susceptibility of the vacuum and the value of the chiral condensate.  The same phenomenon is expected in Minkowski 4d~\cite{our4d}, and we have provided several arguments in this work in support of the idea that even an expanding universe will exhibit the very same properties.

\end{document}